\documentclass[conference]{IEEEtran}
\IEEEoverridecommandlockouts
\usepackage{cite}
\usepackage{amsmath,amssymb,amsfonts}
\usepackage{algorithmic}
\usepackage[ruled,vlined]{algorithm2e}
\usepackage{multirow}
\usepackage{subfigure}
\usepackage{graphicx}
\usepackage{textcomp}
\usepackage{xcolor}
\def\BibTeX{{\rm B\kern-.05em{\sc i\kern-.025em b}\kern-.08em
    T\kern-.1667em\lower.7ex\hbox{E}\kern-.125emX}}
\begin{document}

\title{Adaptive Network Embedding with Arbitrary Multiple Information Sources in Attributed Graphs}

\author{\IEEEauthorblockN{Meng Qin}
\IEEEauthorblockA{\textit{Independent Research} \\
mengqin\_az@foxmail.com}
}

\maketitle

\begin{abstract}
Graph representation learning (a.k.a. network embedding) is a significant topic of network analysis, due to its effectiveness to support various graph inference tasks. In this paper, we study the representation learning with multiple information sources in attributed graphs. Recent studies usually focus on several specific sources (e.g., high-order proximity and node attributes) but few of them can be extended to incorporate other available sources not specified. In addition, most existing methods assume that all the integrated sources share consistent latent features but may ignore the possible inconsistency among them, lacking the required robustness. To address these issues, we propose a novel adaptive hybrid graph representation (AHGR) method from a view of graph reweighting, where each information source is formulated as a corresponding auxiliary graph, enabling AHGR to integrate arbitrary available information sources. Moreover, a new transition relation among the reweighted graphs is then introduced to perceive and resist the possible inconsistency among multiple sources, enhancing the robustness of AHGR. We verify the effectiveness of AHGR on a series of synthetic and real attributed graphs, where it presents superior performance over other baselines.
\end{abstract}


\section{Introduction}
For various complex systems (e.g., social and communication networks), graph is an abstraction describing systems' entities and their relations with sets of nodes and edges. Graph representation learning (a.k.a. network embedding) has emerged as a significant topic in network analysis. Given a graph, it aims to encode the high-dimensional entity relations into a low-dimensional vector representation with the graph's major properties preserved, which can further support various graph inference tasks (e.g., community detection and link prediction) \cite{qintowards,lei2018adaptive,qin2022temporal,qin2023high,qin2023semantic}.

For most network embedding techniques, graph structure (i.e., topology) is the directly available information source. Typical structure information includes the first- and second-order proximities (e.g., similarity between node pairs' neighbors) \cite{Tang2015LINE,Wang2016Structural}, which can be further extended to graph structures with $k$-steps random walk (i.e., high-order proximities) \cite{Perozzi2014DeepWalk,Aditya2016node2vec,Cao2015GraRep}. In addition to the aforementioned microscopic
structure (i.e., different orders of proximities), the mesoscopic community structure is another potential source of graph topology \cite{Liang2016Modularity,Wang2017Community}, which can help to explore the deep organization and function of a graph \cite{Qin2018Adaptive,qin2021dual}. Furthermore, graph semantic (e.g., node attributes) is also a significant source available in attributed graphs \cite{Cheng2015Network,Huang2017Accelerated,Bandyopadhyay2018FSCNMF}, with complementary knowledge beyond graph structures \cite{qin2019towards,li2019identifying}.

It is strongly believed that the incorporation of multiple information sources can potentially enhance the ability of the learned representations to support the downstream tasks, since different sources may reflect distinct aspects of a graph. A series of prior studies try to integrate different sources and have achieved improved performance \cite{Wang2017Community,Cheng2015Network,Huang2017Accelerated,Bandyopadhyay2018FSCNMF}. However, they may suffer from the following limitations.

First, most existing hybrid embedding methods only focus on several specific information sources of a graph but few of them can be extended to integrate other available sources not specified. For instance, several approaches with graph structures and semantic tend to integrate high-order proximities and node attributes \cite{Cheng2015Network,Huang2017Accelerated,Bandyopadhyay2018FSCNMF} but they cannot further explore other sources (e.g., community structure), which may potentially result in better performance for downstream inference tasks.

Second, most existing approaches with multiple information sources are based on the assumption that all the sources share consistent characteristics, ignoring the possible \textit{inconsistency} among the heterogeneous sources. In fact, such \textit{inconsistency} is common in real graphs and may affect the performance of downstream applications (e.g., community detection) \cite{Qin2018Adaptive,qin2021dual}. For instance, \cite{Chakraborty2015Nonnegative} indicated that the social relationships in Twitter can reflect the user group (i.e., ground-truth of the downstream application) more directly than the diverse user-generated content (i.e., node attributes). Therefore, the simple incorporation of attributes may unexpectedly bring inconsistent features to the learned representations.

To alleviate the aforementioned limitations, we proposed a novel adaptive hybrid graph representation (AHGR) method from an alternative view of graph reweighting, which can potentially integrate arbitrary information sources with the consideration of the possible \textit{inconsistency}.

\begin{figure*}[tb]
    \centering
    \includegraphics[width=0.70\linewidth]{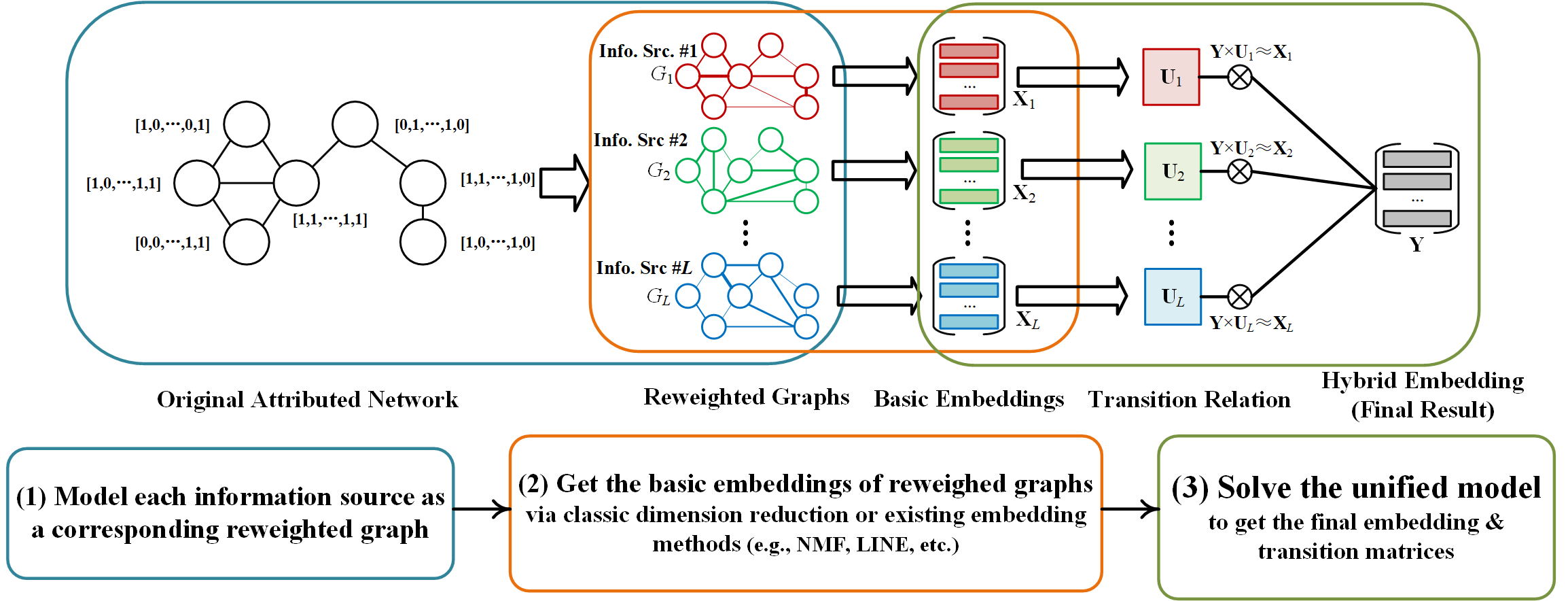}
    \caption{Overview of our AHGR method, which adaptively integrates multiple information sources with the consideration of the \textit{inconsistency effect} in 3 steps.}
    \label{Sketch}
    \vspace{-0.5cm}
\end{figure*}

Fig.~\ref{Sketch} gives a high-level overview of AHGR. Concretely, we first abstract an arbitrary information source as an auxiliary weighted graph, with key characteristics encoded in the weighted topology. For each auxiliary weighted graph, we can derive a corresponding basic low-dimensional representation, defined as the basic embedding, by utilizing a certain dimension reduction technique (e.g., non-negative matrix factorization (NMF) \cite{Lee1999Learning}) or existing embedding method (e.g., LINE \cite{Tang2015LINE}). An NMF-based unified model is then introduced to formulate the \textit{consistency} among multiple sources as a specific \textit{transition relation} from an independent representation to each basic embedding. By solving the unified model, the \textit{transition relation} can adaptively adjust the effect of each basic embedding according to the perceived \textit{inconsistency degree} while the introduced independent representation is considered as the final embedding result.

Our main contributions can be summarized as follows.
\begin{itemize}
    \item We adopt an alternative graph reweighting scheme to formulate the network embedding with multiple information sources, enabling AHGR to integrate arbitrary sources.
    \item To resist the possible \textit{inconsistency} among multiple information sources, we introduce a novel NMF-based transition relation, enhancing the model's robustness.
    \item We develop a novel \textit{consistency indicator} based on the transition relation to quantitatively measure a certain source's \textit{consistency degree}.
    \item We evaluate AHGR's effectiveness on a series of synthetic and real attributed graphs, where it shows superior performance and robustness over other baselines.
\end{itemize}
In the rest of this paper, we introduce the related work in Section~\ref{Rel-work} and give the formal problem statements of graph reweighting and network embedding in Section~\ref{Pro-Def}. Section~\ref{Meth} elaborates on the optimization objective and solving strategy of AHGR. Experiments are described in Section~\ref{Exp}, including the preliminary analysis on synthetic graphs and the performance evaluation on real graphs. Section~\ref{Con} concludes this paper and indicates our future work.

\section{Related Work}\label{Rel-work}
\subsection{Graph Representation Learning}
In the past few years, a series of techniques have been developed for graph representation learning (network embedding). Some related overviews can be found in \cite{Hamilton2017Representation,Cui2017A}.

For most network embedding approaches, graph topology is the directly available source. For instance, \cite{Tang2015LINE} developed a fast embedding methods based on the observed edges (i.e., first-order proximity) and similarity between each node pair's neighbors (i.e., second-order proximity). \cite{Wang2016Structural} further explored the deep non-linear characteristic hidden in the first- and second-order proximities via auto-encoder. Moreover, \cite{Perozzi2014DeepWalk} and \cite{Aditya2016node2vec} extended proximities into higher orders from the view of truncated random walks. Based on matrix factorization (MF), \cite{Qiu2018Network} generalized a series of node-proximity-based methods (e.g., LINE \cite{Tang2015LINE}, DeepWalk \cite{Perozzi2014DeepWalk}, node2vec \cite{Aditya2016node2vec}, etc.) into a unified framework, while \cite{Zhang2018Arbitrary} introduced a unified singular value decomposition (SVD) model for arbitrary-order proximities.

In addition to proximities with microscopic structural information, community structure is another available source revealing the mesoscopic topology. \cite{Liang2016Modularity} reformulated a classic MF-based objective that encodes community membership into an auto-encoder-based nonlinear model while \cite{Wang2017Community} proposed a community-preserved embedding method via NMF.

Furthermore, graph semantic is also a significant source with complementary information beyond topology. For example, methods proposed in \cite{Cheng2015Network} and \cite{Huang2017Accelerated} combined node proximities and attributes via two MF objectives. \cite{Cao2018Incorporating} integrated the community structures and node attributes based on a hybrid auto-encoder.

However, most existing methods only focus on several specific information sources but cannot integrate other available sources not specified. Although some hybrid approaches \cite{Jin2018Integrative} provided naive extended options for the integration of additional information by simply concatenating the multiple inputs, they still inherently ignored the \textit{inconsistency} among the incorporated sources. How to adaptively integrated multiple information sources with the consideration of the possible \textit{inconsistency} is the main focus of this study.

\subsection{The Inconsistency Effect in Attributed Graphs}
The \textit{inconsistency} among multiple information sources has been validated by several prior studies regarding some concrete inference tasks. In \cite{Newman2015Structure}, the authors demonstrated that different types of meta-data (i.e., node attributes) may have distinct effects (i.e., performance improvement or even degradation) to the specific inference application. Focusing on the community detection task, \cite{Qin2018Adaptive} also verified the intrinsic \textit{inconsistency correlation} between the first-order topology and node attribute via a well-designed pre-experiment, where the source of attributes may even damage the learned community structure (i.e., degrade the performance for the inference task) when the \textit{inconsistency effect} occurs, although it can indeed improve the performance in some cases. Moreover, \cite{He2017Joint} and \cite{Jin2018Robust} also developed two node clustering models with much stronger robustness by further considering the possible \textit{inconsistency} between graph topology and attributes.

Nevertheless, the aforementioned studies only focus on one concrete inference task (e.g., community detection). Few of them have considered the \textit{inconsistency} problem in network embedding. To extend the \textit{inconsistency effect} from concrete task to network embedding is the primary goal of this paper.

\section{Problem Statements}\label{Pro-Def}
In this study, we consider graph representation learning (a.k.a. network embedding) in undirected unweighted graphs with node attributes. In general, an attributed graph can be described as a 4-tuple $G=(V,E,A,F)$, where $V=\{ {v_1}, \cdots ,{v_N}\}$ is the node set; $E=\{ ({v_i},{v_j})\left| {{v_i},{v_j} \in V} \right.\}$ is the edge set; $A = \{ {a_1}, \cdots ,{a_M}\}$ is the attribute set; $F = \{ f({v_1}), \cdots ,f({v_N})\}$ is the map from $V$ to $A$ with $f\left( {{v_i}} \right) \subseteq A$ as ${{v_i}}$'s attribute set. Assume that there are $N$ nodes with totally $M$ node attributes and the number of the available information sources is $L$. We use an adjacency matrix ${\bf{A}} \in {\Re ^{N \times N}}$ to describe $G$'s topology, where ${{\bf{A}}_{ij}}={{\bf{A}}_{ji}}=1$ when there is an edge between node pair $({v_i},{v_j})$ and ${{\bf{A}}_{ij}}={{\bf{A}}_{ji}}=0$ otherwise. Graph attributes can be described by a node attribute matrix ${\bf{C}} \in {\Re ^{N \times M}}$, where ${{\bf{C}}_{iw}}=1$ if attribute ${a_w}$ is in node ${v_i}$'s attribute set and ${{\bf{C}}_{iw}}=0$ otherwise.

\textbf{Graph Reweighting Scheme.} We adopt an alternative graph reweighting scheme to encode an arbitrary information source (e.g., high-order proximities, community structures, and attributes). Given a source $l$ (notated as ${I_l}$), one can derive an auxiliary weighted graph ${G_l} = (V,{E_l})$ with ${E_l} = \{ {W_l}({v_i},{v_j})\left| {{v_i},{v_j} \in V} \right.\}$ as the set of weighted edges. Especially, ${G_l}$ can be considered as the reweighting of $G$'s topology with ${I_l}$'s key properties encoded in ${E_l}$. The weighted topology can be described by another adjacency matrix ${{\bf{M}}_l}\in {\Re ^{N \times N}}$ with ${({{\bf{M}}_l})_{ij}} = {({{\bf{M}}_l})_{ji}}$ as $({v_i},{v_j})$'s weight.

\textbf{Graph Representation Learning.} Given the set of auxiliary weighted graphs $\{ {G_1},\cdots ,{G_L}\}$ w.r.t. $\{ {I_1}, \cdots ,{I_L}\}$, we formulate graph representation learning as
\begin{equation}
    {\bf{Y}} = f({{\bf{M}}_1},{{\bf{M}}_2}, \cdots ,{{\bf{M}}_L}),
\end{equation}
such that the primary properties of $\{ {{\bf{M}}_1}, \cdots ,{{\bf{M}}_L}\}$ are preserved in ${\bf{Y}} \in {\Re ^{N \times K}}$. Namely, $f( \cdot )$ maps each node ${v_i}$ to a $K$-dimensional vector ${{\bf{Y}}_{i,:}}$ (with $K \ll \min \{ N,M\}$), where nodes with similar properties (e.g., clustering membership and attributes) should have similar representations.

\section{Methodology}\label{Meth}
\subsection{The Graph Reweighting Scheme}
As a demonstration, we consider the integration of high-order proximities, community structures, and node attributes, which are typical and commonly-used information sources in attributed graphs, but our graph reweighting scheme is not limited to these sources.

\textbf{High-Order Proximities.} The high-order proximities regarding graph topology can be encoded from the view of random walk. For the topology described by an adjacency matrix ${\bf{A}}$, we use the $h$-th power of ${\bf{A}}$ to represent the auxiliary weighted graph w.r.t. $h$-step proximity, where ${({{\bf{A}}^h})_{ij}} = {({{\bf{A}}^h})_{ji}}$ is the number of paths between $({v_i},{v_j})$ with $h$-steps. Since different proximities may reveal distinct structural information of a graph \cite{Cao2015GraRep}, we separately treat each step of proximity as an independent source. Namely, $\{{{\bf{A}}^1}, \cdots ,{\bf{A}}{}^h\}$ are treated as adjacency matrices of $h$ auxiliary weighted graphs regarding high-order proximities.

\textbf{Community Structures.} We use the modularity matrix ${\bf{B}} \in {\Re ^{N \times N}}$ to encode the community structures of a graph, where
\begin{equation}
    {{\bf{B}}_{ij}} = {{\bf{B}}_{ji}} = {{\bf{A}}_{ij}} - {{{d_i}{d_j}} \mathord{\left/ {\vphantom {{{d_i}{d_j}} {(2e)}}} \right. \kern-\nulldelimiterspace} {(2e)}},
\end{equation}
with ${d_i} = \sum\nolimits_{j = 1}^N {{{\bf{A}}_{ij}}}$ as node ${v_i}$'s degree and $e$ as the number of edges. Concretely, $\bf{B}$ measures the difference between the exact number of edges and expected number of such edges over all pairs of nodes. From the view of graph-cut, ${{\bf{B}}_{ij}}$ (${{\bf{B}}_{ji}}$) with larger value means the corresponding edge $({v_i},{v_j} )$ is more likely to be preserved in a certain community but not to be cut for the commuity partitioning, thus encoding the key properties of community structures. We use $\bf{B}$ as the adjacency matrix of the corresponding auxiliary weighted graph.

\textbf{Node Attributes.} To construct the auxiliary weighted graph w.r.t. node attributes, we introduce an attribute similarity matrix ${\bf{S}} \in {\Re ^{N \times N}}$ based on the attribute matrix ${\bf{C}}$, where 
\begin{equation}
    {{\bf{S}}_{ij}} = {{\bf{S}}_{ji}} = {{({{\bf{C}}_{i,:}} \cdot {\bf{C}}_{j,:}^T)} \mathord{\left/ {\vphantom {{({{\bf{C}}_{i,:}} \cdot {\bf{C}}_{j,:}^T)} {(\left| {{{\bf{C}}_{i,:}}} \right|\left| {{{\bf{C}}_{j,:}}} \right|)}}} \right. \kern-\nulldelimiterspace} {(\left| {{{\bf{C}}_{i,:}}} \right|\left| {{{\bf{C}}_{j,:}}} \right|)}}.
\end{equation}
Namely, we construct a weighted graph described by $\bf{S}$ that encodes the semantic similarity between each node pair. In $\bf{S}$, larger ${{\bf{S}}_{ij}}$ (${{\bf{S}}_{ji}}$) indicates ${v_i}$ and ${v_j}$ are more similar in terms of attributes, which should also be preserved in the final embedding result.

When obtaining the adjacency matrices $\{ {{\bf{M}}_1}, \cdots ,{{\bf{M}}_L}\}$ w.r.t. multiple auxiliary weighted graphs $\{{G_1}, \cdots ,{G_L}\}$, there may remain magnitude differences between their weighted topology, unfairly affecting the optimization of AHGR. We use the Z-score and Max-Min normalization in sequence to rescale edge weights to $[0, 1]$. Given a vector/matrix input $\bf{z}$ (e.g., $G_l$'s adjacency matrix ${{\bf{M}}_l}$), the two normalization processes (notated as ZNorm and MNorm) are defined as
\begin{equation}
    {\mathop{\rm ZNorm}\nolimits} ({{\bf{z}}_i}) = \frac{{{{\bf{z}}_i} - \mu }}{\sigma },{\rm{ }}{\mathop{\rm MNorm}\nolimits} ({{\bf{z}}_i}) = \frac{{{{\bf{z}}_i} - {{\bf{z}}_{\min }}}}{{{{\bf{z}}_{\max }} - {{\bf{z}}_{\min }}}},
\end{equation}
where $\mu$ and $\sigma$ are the mean and standard deviation of $\bf{z}$; ${{{\bf{z}}_{\min }}}$ and ${{{\bf{z}}_{\max }}}$ are the maximum and minimum elements in $\bf{z}$.

\subsection{The Basic Embedding}
In AHGR, each reweighted graph ${G_l}$ has its basic embedding (notated as ${{\bf{X}}_l}$), which can be simply derived via a classic dimension reduction technique (e.g., SVD and NMF). We adopt NMF \cite{Lee1999Learning} as the recommended method. Given the adjacency matrix ${{\bf{M}}_l}$ w.r.t. ${I_l}$'s weighted graph, we obtain the basic embedding ${{\bf{X}}_l} \in {\Re ^{N \times K}}$ by solving a symmetrical NMF (SNMF) problem with $l_2$-regularization:
\begin{equation}
    \mathop {\arg \min }\limits_{{{\bf{X}}_l} \ge 0} \frac{1}{2}\left\| {{{\bf{M}}_l} - {{\bf{X}}_l}{\bf{X}}_l^T} \right\|_F^2 + {\lambda _l}\left\| {{{\bf{X}}_l}} \right\|_F^2,
\end{equation}
where ${\lambda_l}$ is used to adjust the regularization term; ${({{\bf{X}}_l})_{i,:}}$ is basic embedding of node ${v_i}$. 

In addition to the classic dimension reduction (e.g., NMF), the basic embedding can also be derived via existing embedding approaches for weighted graphs (e.g., LINE \cite{Tang2015LINE} and SDNE \cite{Wang2016Structural}). We adopt LINE as another method to derive basic embeddings and compare its effectiveness with NMF.

Note that the magnitude difference may still exist among the representation vectors of different nodes. The embedding learned by an embedding method (e.g., LINE) may also not be non-negative, which cannot be directly used by the NMF-based unified model of AHGR. To avoid these issues, we conduct the normalization described in (4) for each row of ${{\bf{X}}_l}$, with the normalized result notated as ${{{\bf{\hat X}}}_l}$.

\subsection{The Unified Model}
For an inference task (e.g., community detection), if there is obvious \textit{inconsistency} among available information $\{{I_1}, \cdots,{I_L}\}$, there should be one or more dominant sources relative to the application's ground-truth (e.g., community membership) with others as irrelevant sources. We introduce an independent non-negative representation ${\bf{Y}} \in {\Re ^{N \times K}}$, which is expected to be in accordance with the application's ground-truth, and reformulate the \textit{inconsistency} among $\{{I_1}, \cdots,{I_L}\}$ as the \textit{consistency relation} between ${\bf{Y}}$ and each basic embedding ${{\bf{X}}_l}$.

Given the normalized basic embedding ${{{\bf{\hat X}}}_l}$ w.r.t. ${I_l}$, we introduce a transition matrix ${{\bf{U}}_l} \in {\Re ^{K \times K}}$ that satisfies $\sum\nolimits_{k = 1}^K {{{\bf{Y}}_{ik}}{{({{\bf{U}}_l})}_{kr}} = {{({{{\bf{\hat X}}}_l})}_{ir}}}$ with ${({{\bf{U}}_l})_{kr}}$ as a corresponding transition weight. Based on such transition relation, we can derive an NMF-based unified model integrating $\{{I_1}, \cdots,{I_L}\}$:
\begin{equation}
    \mathop {\arg \min }\limits_{{\bf{Y}} \ge 0,{{\bf{U}}_l} \ge 0} \sum\limits_{l = 1}^L {(\left\| {{\bf{Y}}{{\bf{U}}_l} - {{{\bf{\hat X}}}_l}} \right\|_F^2 + {\delta _l}\left\| {{{\bf{U}}_l}} \right\|_F^2)}  + \delta \left\| {\bf{Y}} \right\|_F^2,
\end{equation}
where $\{ {\delta _1}, \cdots ,{\delta _L},\delta \}$ are parameters to control the regular terms; $\bf{Y}$ is used as the final embedding result.

In the unified model (6), $\bf{Y}$ is shared by all the information sources, while each source ${I_l}$ has its private transition matrix ${{\bf{U}}_l}$. The \textit{consistency} between ${\bf{Y}}$ and ${{\bf{X}}_l}$ can be encoded in ${{\bf{U}}_l}$. On the one hand, if there are one or more dominant elements with larger transition weights than others in each column of ${{\bf{U}}_l}$ (i.e., an explicit transition from $\bf{Y}$ to ${{\bf{X}}_l}$), ${I_l}$ should be a dominant information source consistent with the application's ground-truth, where ${I_l}$'s properties (encoded by ${{\bf{X}}_l}$) can be fully preserved in $\bf{Y}$ during the optimization. On the other hand, if all the elements in a column of ${{\bf{U}}_l}$ have close values (i.e., an indistinct transition from $\bf{Y}$ to ${{\bf{X}}_l}$), ${I_l}$ is inconsistent with $\bf{Y}$, where ${I_l}$'s effect can be adaptively controlled w.r.t. the perceived \textit{inconsistency degree} in the joint optimization. This property of (6) can also be used to quantitatively measure each source's \textit{consistency degree}. We introduce a novel \textit{consistency indicator} based on the following \textbf{Theorem 1}.

\textbf{Theorem 1.} If we normalize each column of ${{\bf{U}}_l}$ by setting ${({{{\bf{\hat U}}}_l})_{kr}} = {{{{({{\bf{U}}_l})}_{kr}}} \mathord{\left/ {\vphantom {{{{({{\bf{U}}_l})}_{kr}}} {\sum\nolimits_{k = 1}^K {{{({{\bf{U}}_l})}_{kr}}} }}} \right. \kern-\nulldelimiterspace} {\sum\nolimits_{k = 1}^K {{{({{\bf{U}}_l})}_{kr}}} }}$, then $\left\| {{{({{{\bf{\hat U}}}_l})}_{:,r}}} \right\|_2^2$'s value is within $[{1 \mathord{\left/ {\vphantom {1 K}} \right. \kern-\nulldelimiterspace} K},1]$.

\textbf{Proof 1.} To proof Theorem 1 is equivalent to obtain the minimum and maximum value of the following objective:
\begin{equation}
    L = \sum\limits_{k = 1}^K {({{{\bf{\hat U}}}_l})_{kr}^2} \mathop {}\limits_{} {\rm{s}}{\rm{.t}}{\rm{.}}\mathop {}\limits_{} \sum\limits_{k = 1}^K {{{({{{\bf{\hat U}}}_l})}_{kr}}}  = 1,\mathop {}\limits_{} 0 \le {({{{\bf{\hat U}}}_l})_{kr}} \le 1.
\end{equation}
According to (7), there exists the inequality that 
\begin{equation}
\resizebox{.91\linewidth}{!}{$
    1 = {[\sum\limits_{k = 1}^K {{{({{{\bf{\hat U}}}_l})}_{kr}}} ]^2} = \left[ \begin{array}{c}
    \sum\nolimits_{k = 1}^K {({{{\bf{\hat U}}}_l})_{kr}^2 + } \\
    2\sum\nolimits_{s < t}^K {{{({{{\bf{\hat U}}}_l})}_{sr}}{{({{{\bf{\hat U}}}_l})}_{st}}} 
    \end{array} \right] \ge \sum\limits_{k = 1}^K {({{{\bf{\hat U}}}_l})_{kr}^2 = L}.
$}
\end{equation}
The equality in (8) holds if and only if ${({{{\bf{\hat U}}}_l})_{sr}}{({{{\bf{\hat U}}}_l})_{tr}} = 0$ ($\forall s,t \in \{ 1, \cdots ,K\}$). Since $\sum\nolimits_{k = 1}^K {{{({{{\bf{\hat U}}}_l})}_{kr}}} = 1$, $L=1$ when only one element in the $r$-th column of ${{{\bf{\hat U}}}_l}$ (e.g., ${({{{\bf{\hat U}}}_l})_{sr}}$) equals to 1 with other elements ${({{{\bf{\hat U}}}_l})_{tr}} = 0$ ($1 \le t \le K$ and $t \ne s$), indicating the most explicit transition encoded in ${{{\bf{\hat U}}}_l}$. To further prove $L$'s maximum value is 1, we first assume that $L$'s maximum value $p$ is larger than 1. Then, we have
\begin{equation}
\resizebox{.91\linewidth}{!}{$
    p = \sum\limits_{k = 1}^K {({{{\bf{\hat U}}}_l})_{kr}^2}  > 1 = \sum\limits_{k = 1}^K {{{({{{\bf{\hat U}}}_l})}_{kr}}}  \Rightarrow \sum\limits_{k = 1}^K {{{({{{\bf{\hat U}}}_l})}_{kr}}[{{({{{\bf{\hat U}}}_l})}_{kr}} - 1]}  > 0.
$}
\end{equation}
Because ${({{{\bf{\hat U}}}_l})_{kr}} \in [0,1]$, we also have ${({{{\bf{\hat U}}}_l})_{kr}}[{({{{\bf{\hat U}}}_l})_{kr}} - 1] \le 0$ for arbitrary $k \in [1,K]$, which contradicts with (9). Hence, the maximum value of $L$ is 1.

To get $L$'s minimum value, we introduce the Lagrangian multiplier $\mu$ to formulate the following objective:
\begin{equation}
    L' = \sum\nolimits_{k = 1}^K {({{{\bf{\hat U}}}_l})_{kr}^2 + \mu [1 - \sum\nolimits_{k = 1}^K {{{({{{\bf{\hat U}}}_l})}_{kr}}} ]}.
\end{equation}
By setting ${{\partial L'} \mathord{\left/ {\vphantom {{\partial L'} {\partial {{({{{\bf{\hat U}}}_l})}_{kr}}}}} \right. \kern-\nulldelimiterspace} {\partial {{({{{\bf{\hat U}}}_l})}_{kr}}}} = 0$, we have $\mu  = 2{({{{\bf{\hat U}}}_l})_{kr}}$. Since ${{K\mu } \mathord{\left/ {\vphantom {{K\mu } 2}} \right. \kern-\nulldelimiterspace} 2} = \sum\nolimits_{k = 1}^K {{{({{{\bf{\hat U}}}_l})}_{kr}}} = 1$, we can obtain ${({{{\bf{\hat U}}}_l})_{kr}} = {\mu  \mathord{\left/ {\vphantom {\mu  2}} \right. \kern-\nulldelimiterspace} 2} = {1 \mathord{\left/ {\vphantom {1 K}} \right. \kern-\nulldelimiterspace} K}$. Hence, $L$ achieves its minimum value $\sum\nolimits_{k = 1}^K {{1 \mathord{\left/ {\vphantom {1 {{K^2}}}} \right. \kern-\nulldelimiterspace} {{K^2}}}}  = {1 \mathord{\left/ {\vphantom {1 K}} \right. \kern-\nulldelimiterspace} K}$ when all the elements in the $r$-th column of ${{{{\bf{\hat U}}}_l}}$ equal to ${1 \mathord{\left/ {\vphantom {1 K}} \right. \kern-\nulldelimiterspace} K}$, which indicates the most indistinct transition relation encoded in ${{{\bf{\hat U}}}_l}$. Finally, the proof has been completed.

Based on \textbf{Theorem 1}, we define the \textit{consistency indicator} ${\rho_l}$ w.r.t. ${I_l}$ as
\begin{equation}
\resizebox{.89\linewidth}{!}{$
    {\rho _l} = \frac{1}{K}\sum\limits_{r = 1}^K {\frac{{\left\| {{{({{{\bf{\hat U}}}_l})}_{:,r}}} \right\|_2^2 - {1 \mathord{\left/
 {\vphantom {1 K}} \right.
 \kern-\nulldelimiterspace} K}}}{{1 - {1 \mathord{\left/
 {\vphantom {1 K}} \right.
 \kern-\nulldelimiterspace} K}}}}  = \frac{1}{K}\sum\limits_{r = 1}^K {\frac{{K\left\| {{{({{{\bf{\hat U}}}_l})}_{:,r}}} \right\|_2^2 - 1}}{{K - 1}}},
$}
\end{equation}
whose value range is rescaled from $[{1 \mathord{\left/ {\vphantom {1 K}} \right. \kern-\nulldelimiterspace} K},1]$ to $[0, 1]$. In particular, ${\rho_l}$ is proportional to the \textit{consistency degree} of ${I_l}$, where larger ${\rho_l}$ indicates a more explicit transition relation encoded in ${{\bf{U}}_l}$ and larger \textit{consistency degree} of ${I_l}$.

In AHGR, adjusting the effects of multiple information sources $\{ {I_1}, \cdots ,{I_L}\}$ is two-fold.
\begin{itemize}
    \item The transition matrices $\{ {{\bf{U}}_1}, \cdots ,{{\bf{U}}_L}\}$ can perceive the \textit{inconsistency} among $\{ {I_1}, \cdots ,{I_l}\}$ and adaptively adjust ${I_l}$'s contribution to the final embedding ${\bf{Y}}$ w.r.t. ${I_l}$'s \textit{consistency degree}.
    \item The hyper-parameter ${\delta _l}$ can adjust the sparsity of ${{\bf{U}}_l}$, by which additional prior knowledge about ${I_l}$ can be incorporated. Concretely, larger ${\delta_l}$ makes ${{\bf{U}}_l}$ less sparse (i.e., less explicit transition and smaller \textit{consistency degree} encoded in ${{\bf{U}}_l}$), since the F-norm term $\left\| {{{\bf{U}}_l}} \right\|_F^2$ has the effect of smoothing.
\end{itemize}

\subsection{Model Optimization}
We adopt the block coordinate descent method \cite{Qi2012Community} to solve the non-convex optimization problem defined in (6). To obtain the solution (notated as $\{ {{\bf{Y}}^*},{\bf{U}}_l^*\}$ with $1 \le l \le L$), we first randomly initialize $\{{\bf{Y}},{{\bf{U}}_l}\}$ and alternately take the following two steps to update them until converge.

(\romannumeral1) \textbf{The ${\bf{Y}}$-Step}: In this step, we update ${\bf{Y}}$ with ${{\bf{U}}_l}$ ($1\le l\le L$) fixed. The updating rule can be derived by solving the following optimization problem only related to ${\bf{Y}}$:
\begin{equation}
    \mathop {\arg \min }\limits_{{\bf{Y}} \ge 0} {{\mathop{\rm O}\nolimits} _{\bf{Y}}}({\bf{Y}}) = \sum\nolimits_{l = 1}^L {\left\| {{\bf{Y}}{{\bf{U}}_l} - {{{\bf{\hat X}}}_l}} \right\|_F^2}  + \delta \left\| {\bf{Y}} \right\|_F^2.
\end{equation}
Subsequently, we can derive the partial derivative w.r.t. ${\bf{Y}}$:
\begin{equation}
    \frac{{\partial {{\mathop{\rm O}\nolimits} _{\bf{Y}}}({\bf{Y}})}}{{\partial {\bf{Y}}}} = 2\sum\nolimits_{l = 1}^L {({\bf{Y}}{{\bf{U}}_l}{\bf{U}}_l^T - {{{\bf{\hat X}}}_l}{\bf{U}}_l^T)}  + 2\delta {\bf{Y}}.
\end{equation}
We use simplified notations ${\left[ . \right]_ + }$ and ${\left[ . \right]_ - }$ to represent terms with positive coefficient and negative coefficients,(i.e., ${\left[ . \right]_ + } = 2\sum\nolimits_{l = 1}^L {{\bf{Y}}{{\bf{U}}_l}{\bf{U}}_l^T}  + 2\delta {\bf{Y}}$ and ${\left[ . \right]_ - } = 2\sum\nolimits_{l = 1}^L {{{{\bf{\hat X}}}_l}{\bf{U}}_l^T}$).

By gradient descent, one can obtain the following addictive updating rule:
\begin{equation}
    {{\bf{Y}}_{ir}} \leftarrow {{\bf{Y}}_{ir}} - {\eta _{ir}}{({[.]_ + } - {[.]_ - })_{ir}},
\end{equation}
with ${\eta _{ir}}$ as the learning rate. We further transform (14) into a multiplicative form by setting ${\eta _{ir}} = {{{{\bf{Y}}_{ir}}} \mathord{\left/ {\vphantom {{{{\bf{Y}}_{ir}}} {{{({{[.]}_ + })}_{ir}}}}} \right. \kern-\nulldelimiterspace} {{{({{[.]}_ + })}_{ir}}}}$:
\begin{equation}
\resizebox{.89\linewidth}{!}{$
    {{\bf{Y}}_{ir}} \leftarrow {{\bf{Y}}_{ir}}\frac{{{{({{[ \cdot ]}_ - })}_{ir}}}}{{{{({{[ \cdot ]}_ + })}_{ir}}}} = {{\bf{Y}}_{ir}}\frac{{{{(\sum\nolimits_{l = 1}^L {{{{\bf{\hat X}}}_l}} {\bf{U}}_l^T)}_{ir}}}}{{{{({\bf{Y}}(\sum\nolimits_{l = 1}^L {{{\bf{U}}_l}{\bf{U}}_l^T}  + \delta {{\bf{I}}_K}))}_{ir}}}},
$}
\end{equation}
where ${{{\bf{I}}_K}}$ represents the $K$-dimensional identity matrix.

The aforementioned solving strategy is effective for most NMF-based problems. \cite{Qi2012Community} has proved that if variables (e.g., ${\bf{Y}}$) are initialized with non-negative values, the non-negative constraint (e.g., ${\bf{Y}} \ge 0$) and the strategy's convergence can be guaranteed during the optimization. We adopt a similar strategy to derive the updating rules of $\{ {{\bf{U}}_1}, \cdots ,{{\bf{U}}_L}\}$.

(\romannumeral2) \textbf{The ${\bf{U}}$-Step}: In this step, we alternatively update ${{\bf{U}}_l}$ ($1\le l\le L$) with other variables fixed. To obtain the updating rule, we extract terms only related to ${{\bf{U}}_l}$ in (6) and formulate the following optimization problem:
\begin{equation}
    \mathop {\arg \min }\limits_{{{\bf{U}}_l} \ge 0} {{\mathop{\rm O}\nolimits} _{\bf{U}}}\left( {{{\bf{U}}_l}} \right) = \left\| {{\bf{Y}}{{\bf{U}}_l} - {{{\bf{\hat X}}}_l}} \right\|_F^2 + {\delta _l}\left\| {{{\bf{U}}_l}} \right\|_F^2.
\end{equation}
Accordingly, one can get the partial derivative w.r.t. ${{\bf{U}}_l}$:
\begin{equation}
    \frac{{\partial {{\mathop{\rm O}\nolimits} _{\bf{U}}}\left( {{{\bf{U}}_l}} \right)}}{{\partial {{\bf{U}}_l}}} = 2\left( {{{\bf{Y}}^T}{\bf{Y}}{{\bf{U}}_l} - {{\bf{Y}}^T}{{{\bf{\hat X}}}_l}} \right) + 2{\delta _l}{{\bf{U}}_l}.
\end{equation}
Similar to the ${\bf{Y}}$-step, we can obtain the updating rule of ${{\bf{U}}_l}$ by partitioning terms in (17) into ${\left[ . \right]_ + }$ and ${\left[ . \right]_ - }$:
\begin{equation}
    {({{\bf{U}}_l})_{kr}} \leftarrow {({{\bf{U}}_l})_{kr}}\frac{{{{({{\bf{Y}}^T}{{{\bf{\hat X}}}_l})}_{kr}}}}{{{{({{\bf{Y}}^T}{\bf{Y}}{{\bf{U}}_l} + {\delta _l}{{\bf{U}}_l})}_{kr}}}}.
\end{equation}

Note that the aforementioned strategy can only ensure the local minimum solution but not the global optimum. To avoid this issue, we conduct the solving procedure multiple times (e.g., 10 times in our experiments) and adopt the solution with minimum converged value of objective function (6) as the final result. Furthermore, we adopt the criterion based on the objective value to determine whether the process has converged. In each iteration, we record the relative error of objective function w.r.t. the previous iteration. If the relative error is smaller than a pre-set threshold (e.g., $10^{-6}$ in our experiments), we determine the process has converged.

Algorithm~\ref{AHGR-Alg} summarizes the overall procedure of AHGR. During the optimization, the complexities to conduct the ${\bf{Y}}$-step and ${\bf{U}}$-step once are $O(N{K^2} + {K^3})$ and $O(LN{K^2} + {K^3})$. Since $K \ll N$, the overall complexity is no more than $O\left( {RLN{K^2}} \right)$ for $R$ iterations.

\begin{algorithm}[tb]
\caption{Adaptive Hybrid Graph Representation}
\label{AHGR-Alg}
\LinesNumbered
\KwIn{$\{ {I_1}, \cdots ,{I_L}\}$}
\KwOut{${\bf{Y}}^*$, $\{ {{\bf{U}}_1^*}, \cdots ,{{\bf{U}}_L^*}\}$, $\{ {\rho _1}, \cdots ,{\rho _L}\}$}
\For{$l$ {\bf{from}} $1$ {\bf{to}} $L$}
{
    construct the reweighted graph ${G_l}$ regarding ${I_l}$\\
    normalize ${G_l}$'s adjacency matrix ${{\bf{M}}_l}$ via (4)\\
    get the basic embedding ${{\bf{X}}_l}$ regarding ${{\bf{M}}_l}$\\
    normalize each row of ${{\bf{X}}_l}$ via (4)\\
}
initialize $\{ {\bf{Y}},{{\bf{U}}_1}, \cdots ,{{\bf{U}}_L}\}$\\
\While{{\bf{not}} converge}
{
    update ${\bf{Y}}$ via (15) //${\bf{Y}}$-Step\\
    \For{$l$ {\bf{from}} $1$ {\bf{to}} $L$}
    {
        update ${{\bf{U}}_l}$ via (18) //${\bf{U}}$-Step\\
    }
}
calculate $\{ {\rho _l}\}$ according to $\{{\bf{U}}_l^*\}$ via (11)
\end{algorithm}

The time to get the basic embedding ${{\bf{X}}_l}$ depends on the embedding method we adopted. For the NMF-based approach defined in (5), the solving strategy is similar to that of (6), where we first initialize ${{\bf{X}}_l}$ via the NNDSVD strategy \cite{Boutsidis2008SVD} and iteratively update its value using the following rule until converge:
\begin{equation}
    {({{\bf{X}}_l})_{ir}} \leftarrow {({{\bf{X}}_l})_{ir}}\frac{{{{({{\bf{M}}_l}{{\bf{X}}_l})}_{ir}}}}{{{{({{\bf{X}}_l}{\bf{X}}_l^T{{\bf{X}}_l} + {\lambda _l}{{\bf{X}}_l})}_{ir}}}}.
\end{equation}
Hence, the complexity of the NMF-based embedding method is $O(R{N^2}K)$ for $R$ iterations.

On the other hand, the complexity of LINE is $O(nK\left| {{E_l}} \right|)$ with $n$ and $\left| {{E_l}} \right|$ as the number of negative samples and edges \cite{Tang2015LINE}. For the details of LINE (e.g., the optimization objective, algorithm, etc.), please refer to \cite{Tang2015LINE}.

Note that AHGR is a typical two-stage optimized unified model but not an edge-to-edge joint optimized method. Although the latter jointly-optimized form may potentially lead to a better result, we still design AHGR as a two-stage method based on the consideration of the computation time and memory usage. Usually, the joint optimization of multiple information sources is time-consuming and memory-consuming, since all the sources $\{ {I_1}, \cdots ,{I_L}\}$ are directly incorporated into a unified model from the original data space with high dimensionality (e.g., $N$ and $M$). In contrast, the two-stage optimization first maps each available source into a latent space with much lower dimension (i.e., $K \ll \min \{ N,M\}$). The derivation of basic embeddings is also highly flexible and can be implemented in parallel, which can effectively reduce the time and memory consumption. Concretely, the method to obtain basic embeddings can be existing mature dimension reduction techniques (e.g., NMF with fast distributed implementation) or  fast embedding approaches (e.g., LINE). One can also simultaneously derive the basic embeddings of multiple sources in parallel. Moreover, the complexity of AHGR’s second step is also much smaller than the joint optimization of some hybrid approaches, since all the information are already mapped into the low-dimensional space.

\section{Experiments}\label{Exp}
\subsection{Analysis on Synthetic Graphs}
To verify the AHGR's robustness under a quantitatively controllable condition, we applied it to a series of synthetic attributed graphs with simulated \textit{inconsistency} among different information sources and adopted node clustering (a.k.a. community detection) as an example application.

The synthetic graphs were generated based on the GN-net \cite{Girvan2002Community}, containing 128 nodes and 128 attributes. For graph topology, we evenly partitioned nodes into 4 topology clusters (i.e., each cluster has $p=32$ node members), where each node has average ${z_{{\rm{in}}}}$ edges connecting to other nodes in the same cluster and average ${z_{{\rm{out}}}}$ edges connecting to those in different clusters, with ${z_{{\rm{in}}}} + {z_{{\rm{out}}}} = 16$. For attributes, we evenly partitioned nodes into 4 attribute clusters (with a one-to-one correspondence to topology clusters) according to each node's 128-dimensional attribute vector (i.e., each cluster has $q=32$ relevant attributes), where each node's attribute vector has average ${h_{{\rm{in}}}}$ elements relevant to the cluster it belongs to and average ${h_{{\rm{out}}}}$ irrelevant elements, with ${h_{{\rm{in}}}} + {h_{{\rm{out}}}} = 16$. 

Based on the aforementioned constraints, one can generate the corresponding adjacency matrix ${\bf{A}}$ and attribute matrix ${\bf{C}}$ to describe graph topology and attributes. Concretely, if node ${v_i}$ belongs to the topology (attribute) cluster $s$, one can set each element in the $i$-th row of the adjacency matrix ${\bf{A}}$ (node attribute matrix ${\bf{C}}$) with the column index range $[(s - 1)p + 1,sp]$ ($[(s - 1)q + 1,sq]$) to be 1 following the probability ${{{z_{{\rm{in}}}}} \mathord{\left/ {\vphantom {{{z_{{\rm{in}}}}} p}} \right. \kern-\nulldelimiterspace} p}$ (${{{h_{{\rm{in}}}}} \mathord{\left/ {\vphantom {{{h_{{\rm{in}}}}} q}} \right. \kern-\nulldelimiterspace} q}$), while set other elements in the same row to be 1 with the probability ${{{z_{{\rm{out}}}}} \mathord{\left/ {\vphantom {{{z_{{\rm{out}}}}} {(3p)}}} \right. \kern-\nulldelimiterspace} {(3p)}}$ (${{{h_{{\rm{out}}}}} \mathord{\left/ {\vphantom {{{h_{{\rm{out}}}}} {(3p)}}} \right. \kern-\nulldelimiterspace} {(3p)}}$).

Since the observable topology and attributes are two typical heterogeneous sources, which has been validated to have the potential inconsistent hidden features by prior research \cite{He2017Joint,Jin2018Robust,Bandyopadhyay2018FSCNMF,Qin2018Adaptive,qin2021dual}, we follow \cite{He2017Joint,Qin2018Adaptive,qin2021dual} to simulate the \textit{inconsistency} between the two sources with four cases corresponding to four types of \textit{inconsistency}:

\textbf{(1)} \textit{attribute has the inconsistent clustering membership};

\textbf{(2)} \textit{topology has the inconsistent clustering membership};

\textbf{(3)} \textit{there exist noises in attribute};

\textbf{(4)} \textit{there exist noises in topology}.

Note that we do not consider $h$-step ($h>2$) proximities and community structures, since they may introduce additional structural information compared to attributes, unfairly affecting the evaluation. In the rest of this section, we use subscripts T and A to denote variables w.r.t. graph topology and attributes, respectively.

In Case (1) and (3) (Case (2) and (4)), topology (attributes) is considered as the dominant information relative to the application's ground-truth (i.e., cluster membership). We quantitatively simulated the four types of \textit{inconsistency} by adjusting some parameters. For Case (1)/(2), we set ${z_{{\rm{out}}}} = {h_{{\rm{out}}}} = 8$ and randomly selected a proportion ${\gamma _{{\rm{inc}}}}$ of nodes to swap their corresponding rows in ${\bf{A}}$/$\bf{C}$. Moreover, we varied ${\gamma _{{\rm{inc}}}}$ from 0 to 1 with the step size of 0.1 to gradually increase the \textit{inconsistency degree} between topology and attributes. For Case (3)/(4), we set ${z_{{\rm{out}}}}=8$/${h_{{\rm{out}}}}=8$ and varied ${h_{{\rm{out}}}}$/${z_{{\rm{out}}}}$ from 0 to 12 with step size of 1, where larger ${z_{{\rm{out}}}}$/${h_{{\rm{out}}}}$ means more noises in topology/attributes.

In our experiments, we used the recommended NMF method (5) to derive basic embeddings of topology and attributes (with ${\lambda _{\rm{T}}} = 5$, ${\lambda _{\rm{A}}} = 1$) and also adopted them as two baselines (notated as NMF-T and NMF-A). To verify AHGR's ability to perceive and resist \textit{inconsistency}, we first set $\delta = {\delta_{\rm{T}}} = {\delta_{\rm{A}}} = 1$ for all the cases, with the corresponding results and \textit{consistency indicators} notated by AHGR(0) and ${\rho_*}(0)$, respectively. To further illustrate the auxiliary effect of introducing the prior knowledge, we set ${\delta _{\rm{A}}}/{\delta _{\rm{T}}} \in \{ 1,1,2,2,3,5,5,10,10,10,10\}$ for ${\gamma _{{\rm{inc}}}} \in \{ 0.0,0.1, \cdots ,1.0\}$ in Case (1)/(2) and ${\delta _{\rm{A}}}({\delta _{\rm{T}}}) \in \{ 1,1,1,1,1,1,1,1,2,2,5,5\}$ for ${h_{{\rm{out}}}}/{z_{{\rm{out}}}} \in \{ 0,1, \cdots ,12\}$ in Case (3)/(4) with $\delta = 1$. The corresponding results and \textit{consistency indicators} are denotated by AHGR(1) and ${\rho_*}(1)$. Moreover, TADW \cite{Cheng2015Network}, an embedding approach that integrates graph topology and attributes but inherently ignores the possible \textit{inconsistency}, was used as another baseline.

\begin{figure}[tb]
\centering
\begin{minipage}{0.48\linewidth}
    \centering
    \subfigure[\scriptsize{Case (1): Evaluation Results}]
    {\includegraphics[width=1.0\textwidth, trim=5 0 30 0, clip]{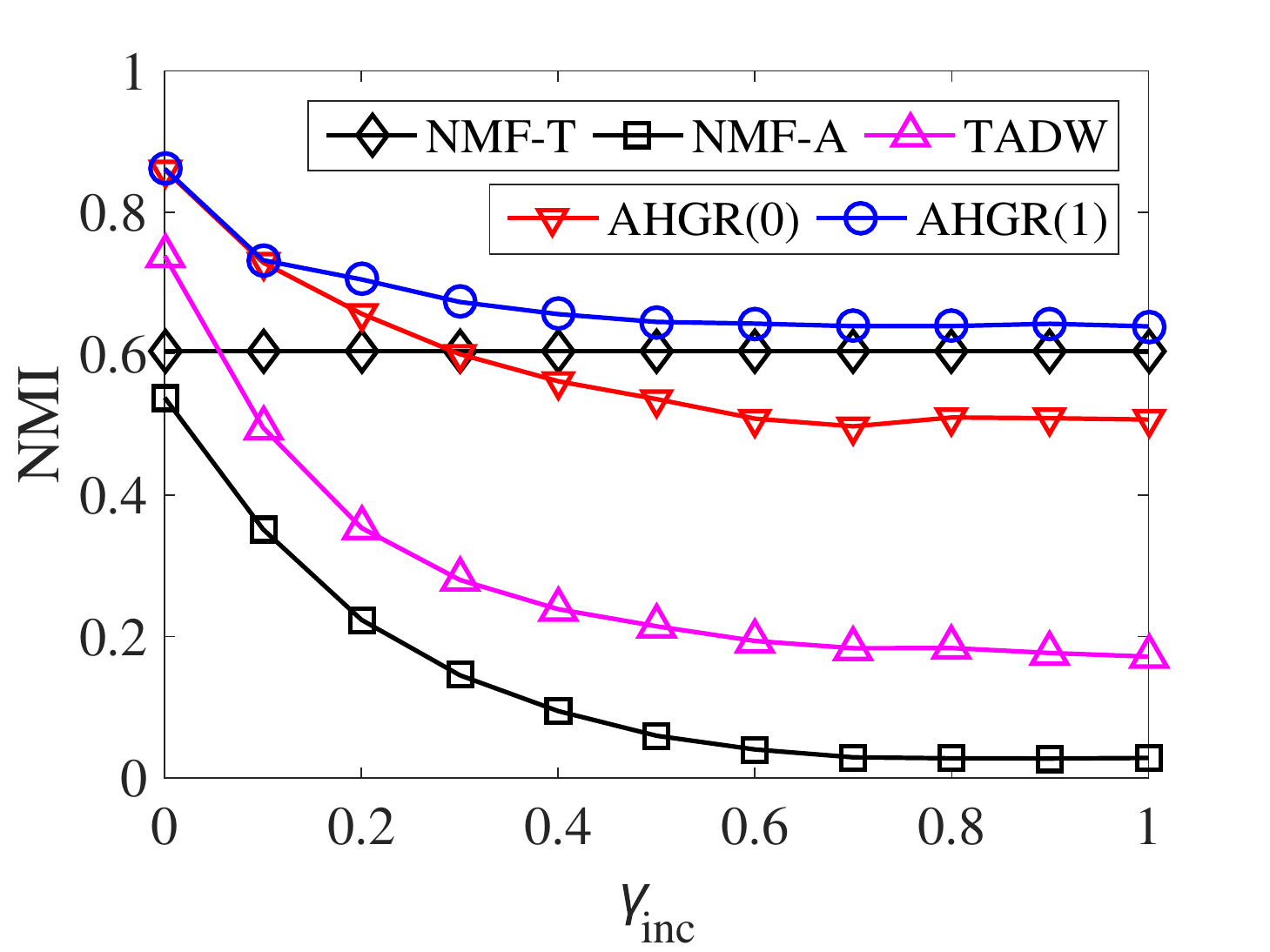}}
\end{minipage}
\begin{minipage}{0.48\linewidth}
    \centering
    \subfigure[\scriptsize{Case (2): Evaluation Results}]
    {\includegraphics[width=1.0\textwidth, trim=5 0 30 0, clip]{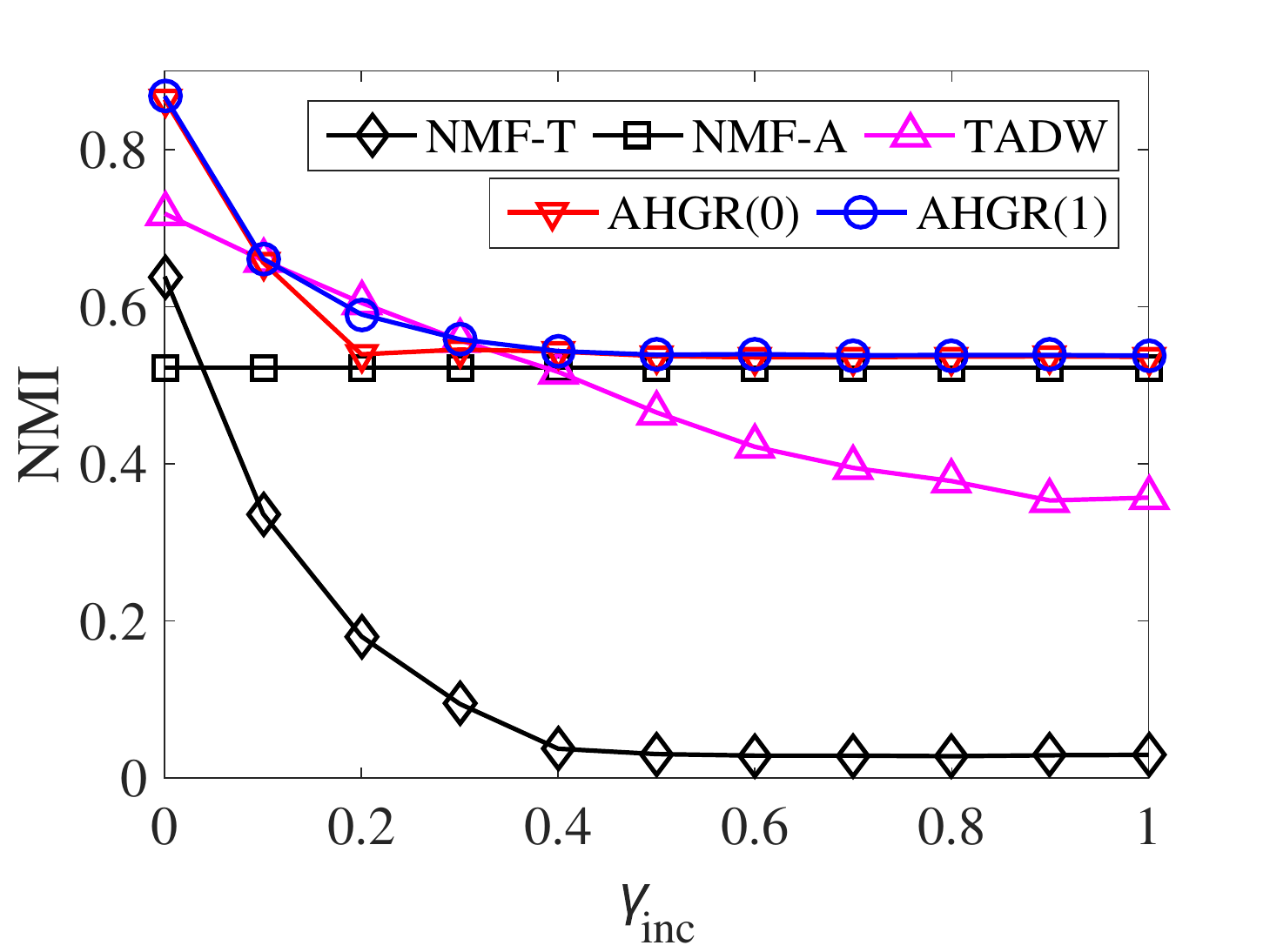}}
\end{minipage}
\begin{minipage}{0.48\linewidth}
    \centering
    \subfigure[\scriptsize{Case (3): Evaluation Results}]
    {\includegraphics[width=1.0\textwidth, trim=5 0 30 0, clip]{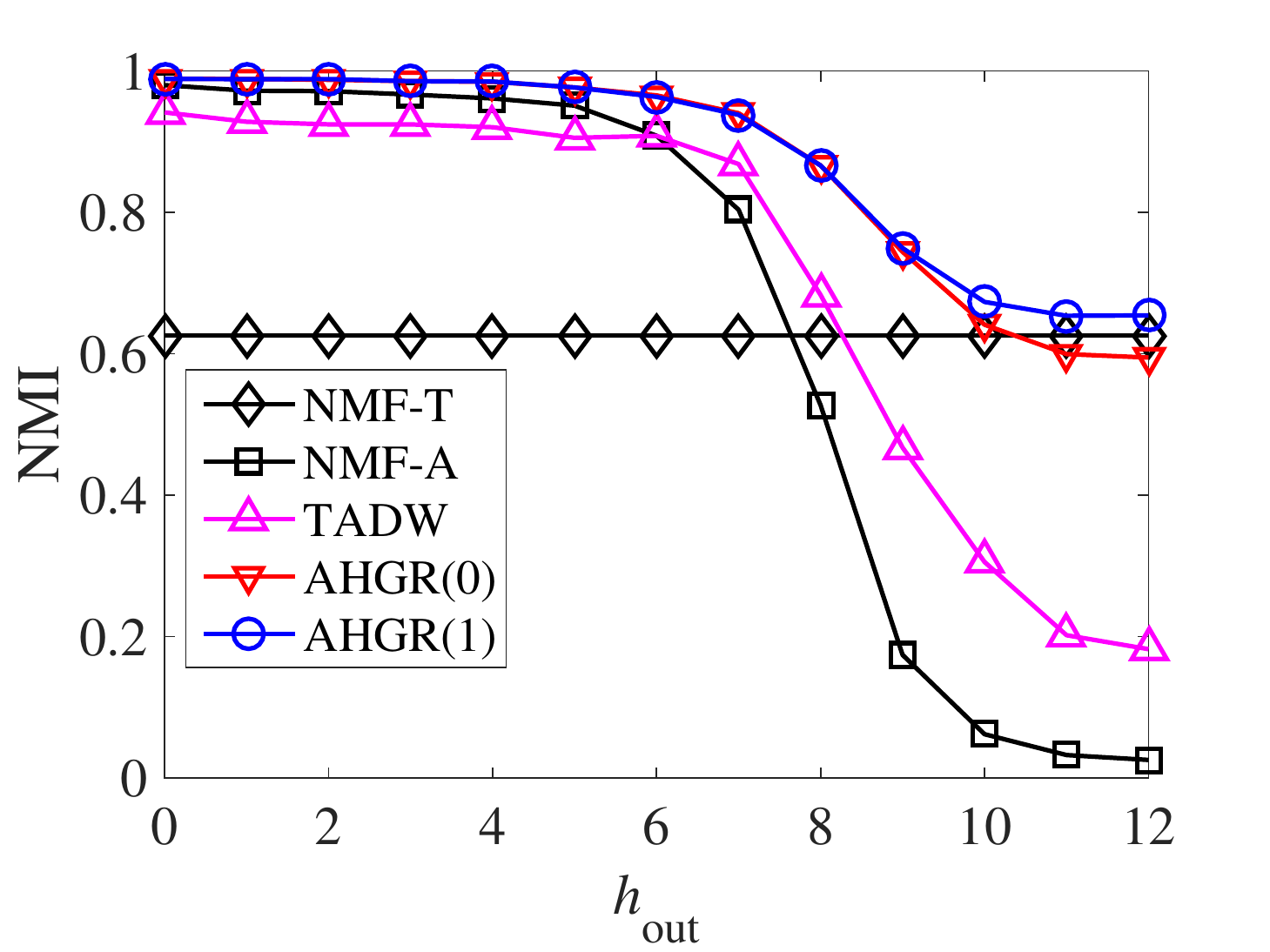}}
\end{minipage}
\begin{minipage}{0.48\linewidth}
    \centering
    \subfigure[\scriptsize{Case (4): Evaluation Results}]
    {\includegraphics[width=1.0\textwidth, trim=5 0 30 0, clip]{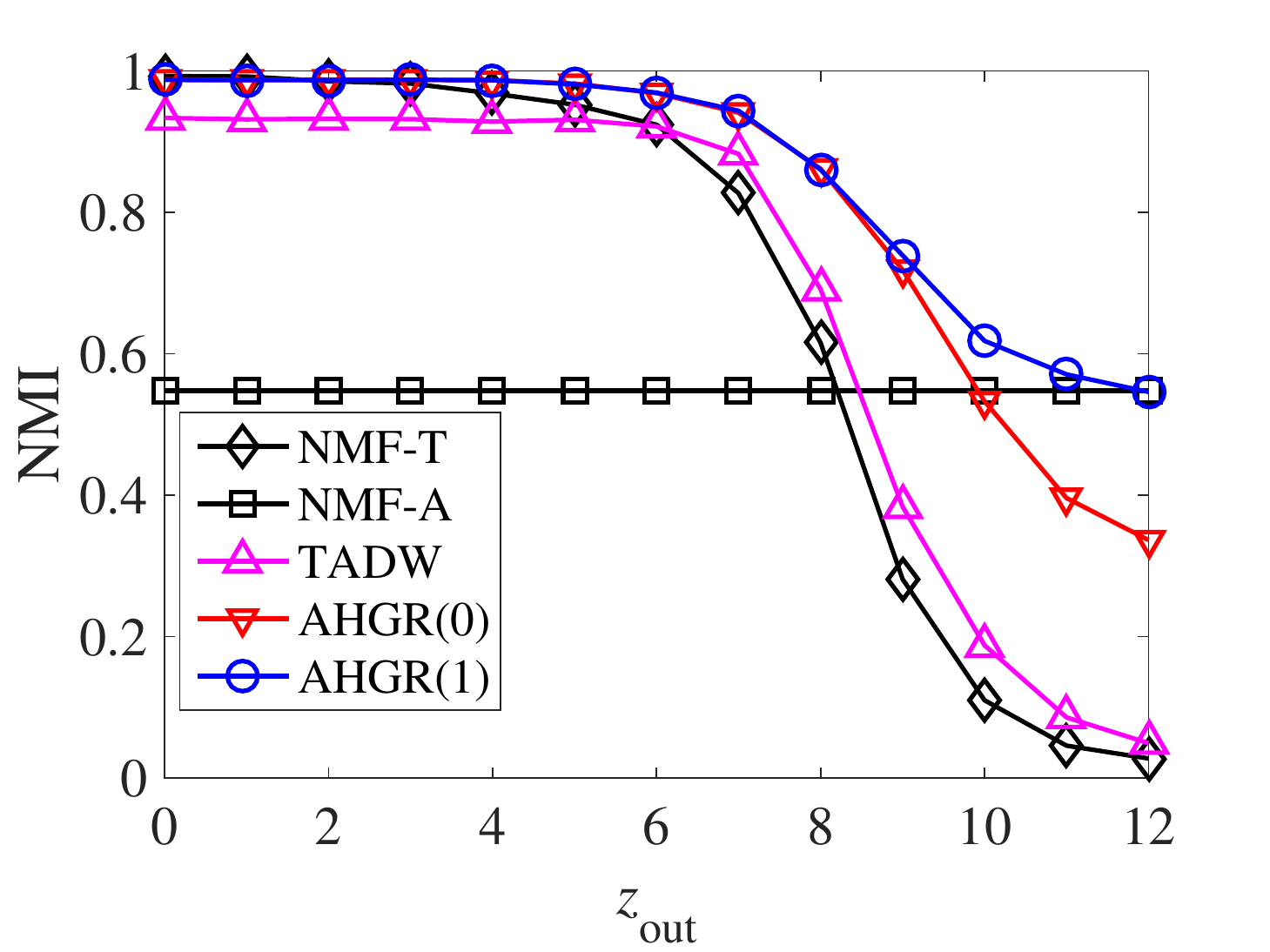}}
\end{minipage}
\begin{minipage}{0.48\linewidth}
    \centering
    \subfigure[\scriptsize{Case (1): \textit{Consistency Indicators}}]
    {\includegraphics[width=1.0\textwidth, trim=5 0 30 0, clip]{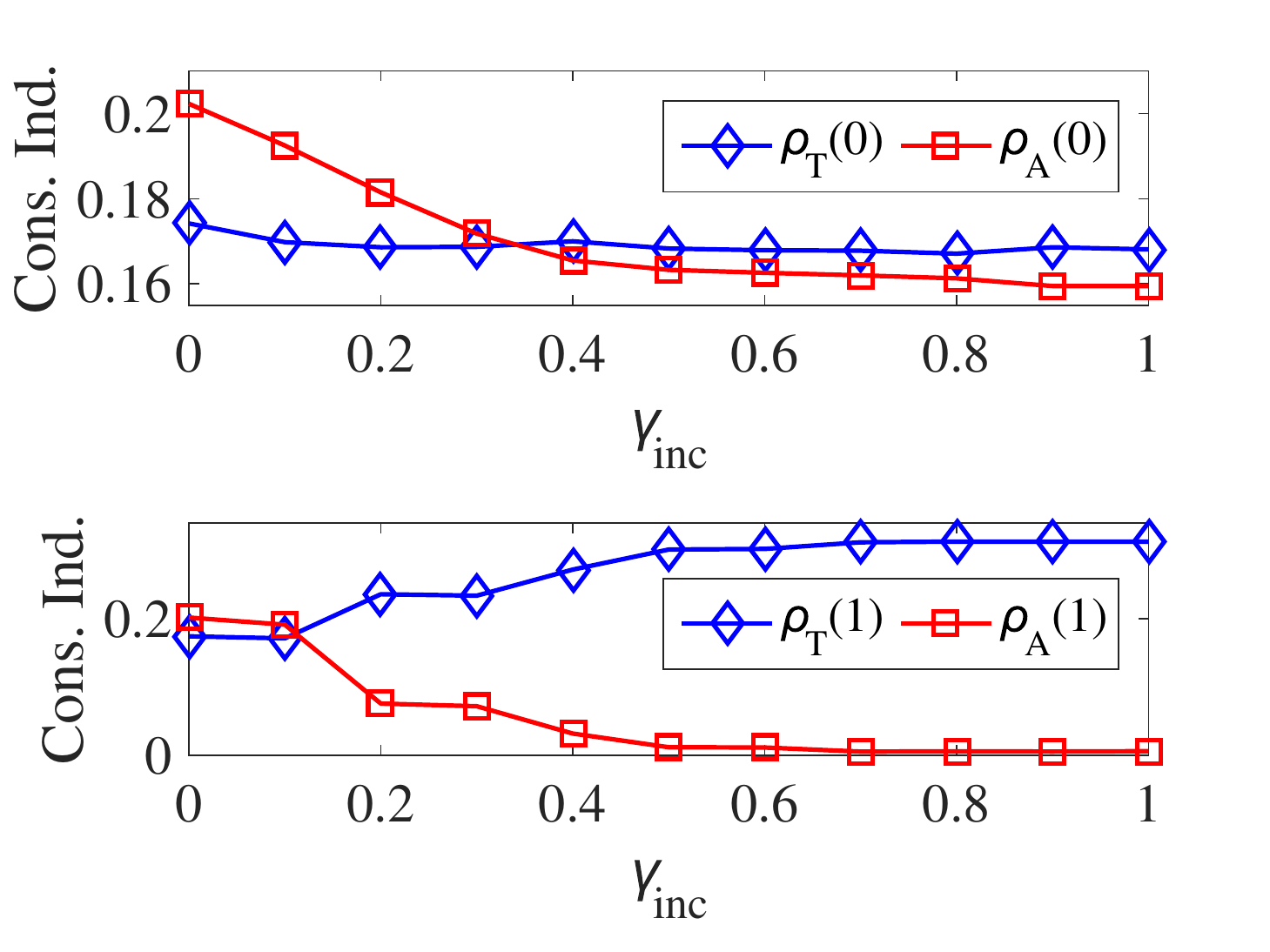}}
\end{minipage}
\begin{minipage}{0.48\linewidth}
    \centering
    \subfigure[\scriptsize{Case (2): \textit{Consistency Indicators}}]
    {\includegraphics[width=1.0\textwidth, trim=5 0 30 0, clip]{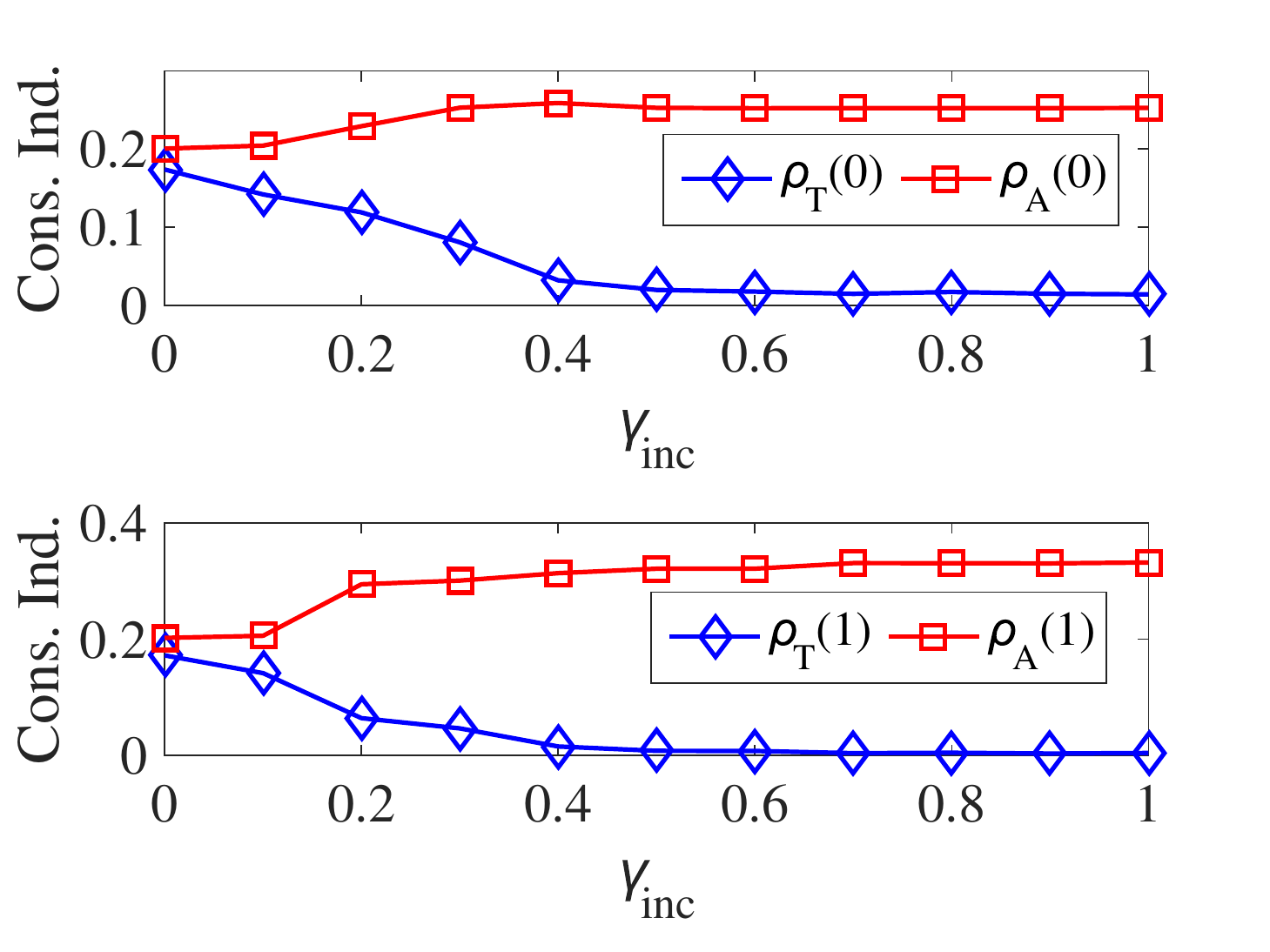}}
\end{minipage}
\begin{minipage}{0.48\linewidth}
    \centering
    \subfigure[\scriptsize{Case (3): \textit{Consistency Indicators}}]
    {\includegraphics[width=1.0\textwidth, trim=5 0 30 0, clip]{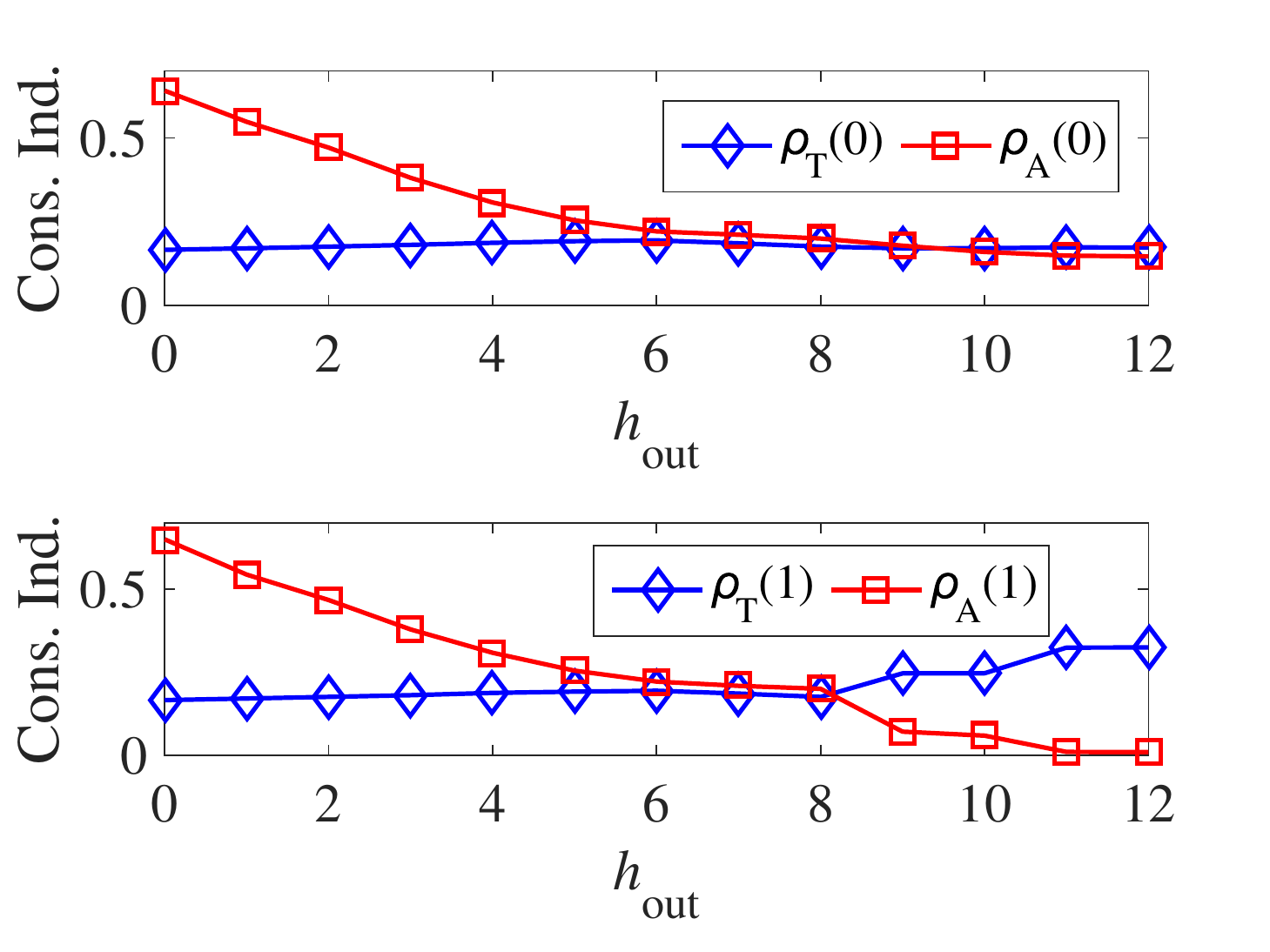}}
\end{minipage}
\begin{minipage}{0.48\linewidth}
    \centering
    \subfigure[\scriptsize{Case (4): \textit{Consistency Indicators}}]
    {\includegraphics[width=1.0\textwidth, trim=5 0 30 0, clip]{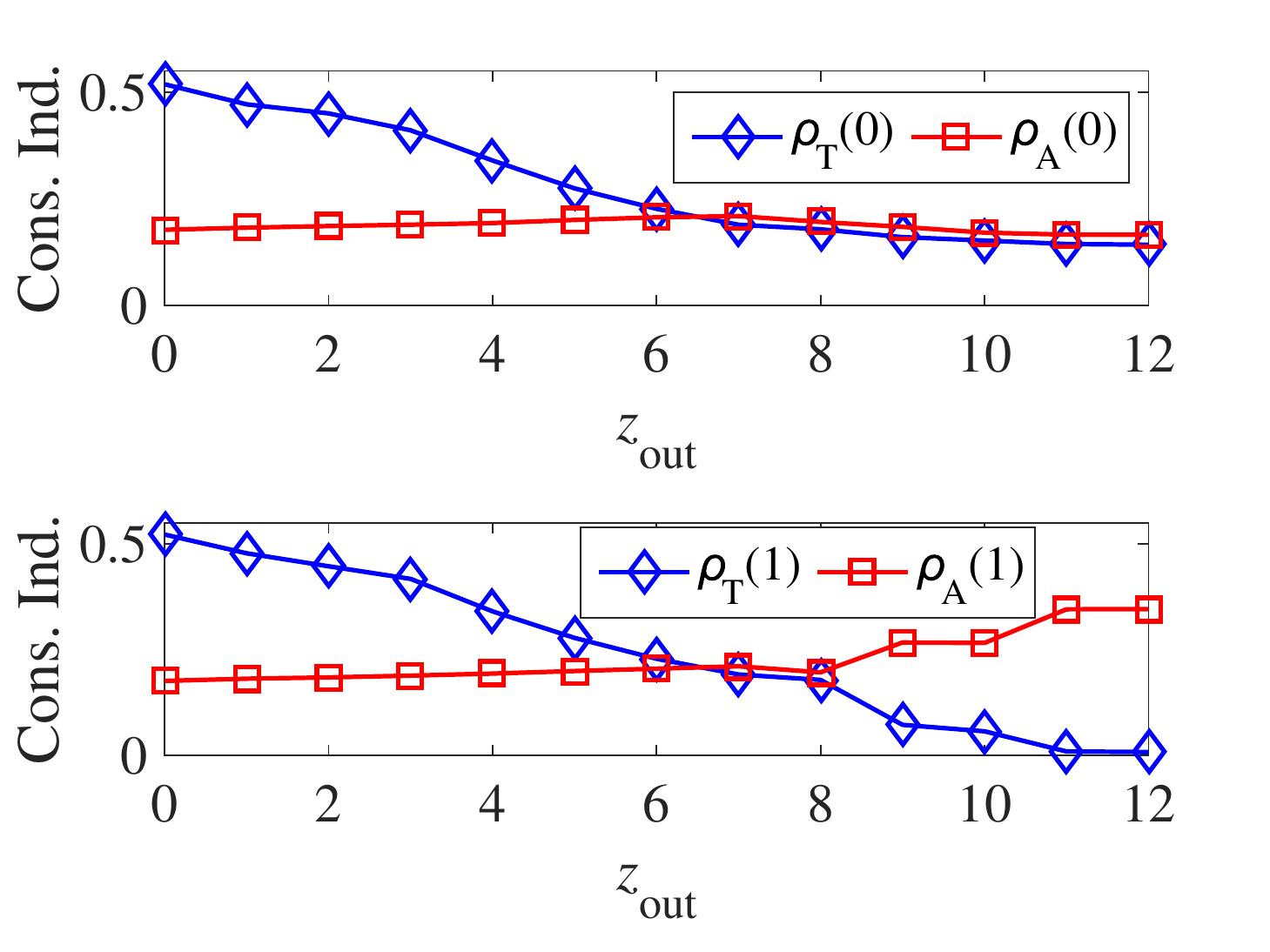}}
\end{minipage}
\caption{Evaluation results and consistency indicator curves of synthetic graph analysis from Case (1) to (4)}\label{Syn-Exp}
\vspace{-0.5cm}
\end{figure}

For each method to be evaluated, we uniformly set $K = 8$ and ran the $K$Means algorithm 100 times on the embedding. The average normalized mutual information (NMI) \cite{Cui2017A} was reported as the quality metric. For AHGR, we also recorded the corresponding \textit{inconsistency indicators}. The overall generating and evaluation process was conducted 50 times, with the average NMIs and \textit{consistency indicators} shown in Fig.~\ref{Syn-Exp}.

In Case (1) (see Fig.~\ref{Syn-Exp} (a) and (e)), when ${\gamma _{{\rm{inc}}}} = 0$, TADW and AHGR outperform the baselines NMF-T and NMF-A, but AHGR still has the better performance. It implies that AHGR can better integrate multiple information sources by fully exploring their intrinsic \textit{consistency} correlations. As ${\gamma _{{\rm{inc}}}}$ increases, the performance of NMF-A and TADW seriously deteriorate with a big gap lower than NMF-T, but AHGR still keeps at a level that near to NMF-T, indicating the powerful robustness of AHGR. In particular, AHGR(1) achieves the NMIs better than NMF-T even when ${\gamma _{{\rm{inc}}}}$ is large, which verifies that the introduction of prior knowledge can further help AHGR to resist the \textit{inconsistency}. Moreover, with the increase of ${\gamma _{{\rm{inc}}}}$, ${\rho _{\rm{A}}}(0)$ and ${\rho _{\rm{A}}}(1)$ consistently decrease and ${\rho _{\rm{T}}}(1)$ also has the obvious increase, which implies the increase of attributes' \textit{inconsistency degree}. Namely, attributes should have less contribution in the joint optimization. Therefore, AHGR's ability to perceive the possible \textit{inconsistency} can be verified. One can reach similar conclusions for other cases.

\begin{figure}[]
\centering
\begin{minipage}{0.48\linewidth}
    \centering
    \subfigure[\scriptsize{${{\bf{U}}_{\rm{T}}}$ (${\gamma} = 0$)}]
    {\includegraphics[width=1.0\textwidth, trim=30 30 20 20, clip]{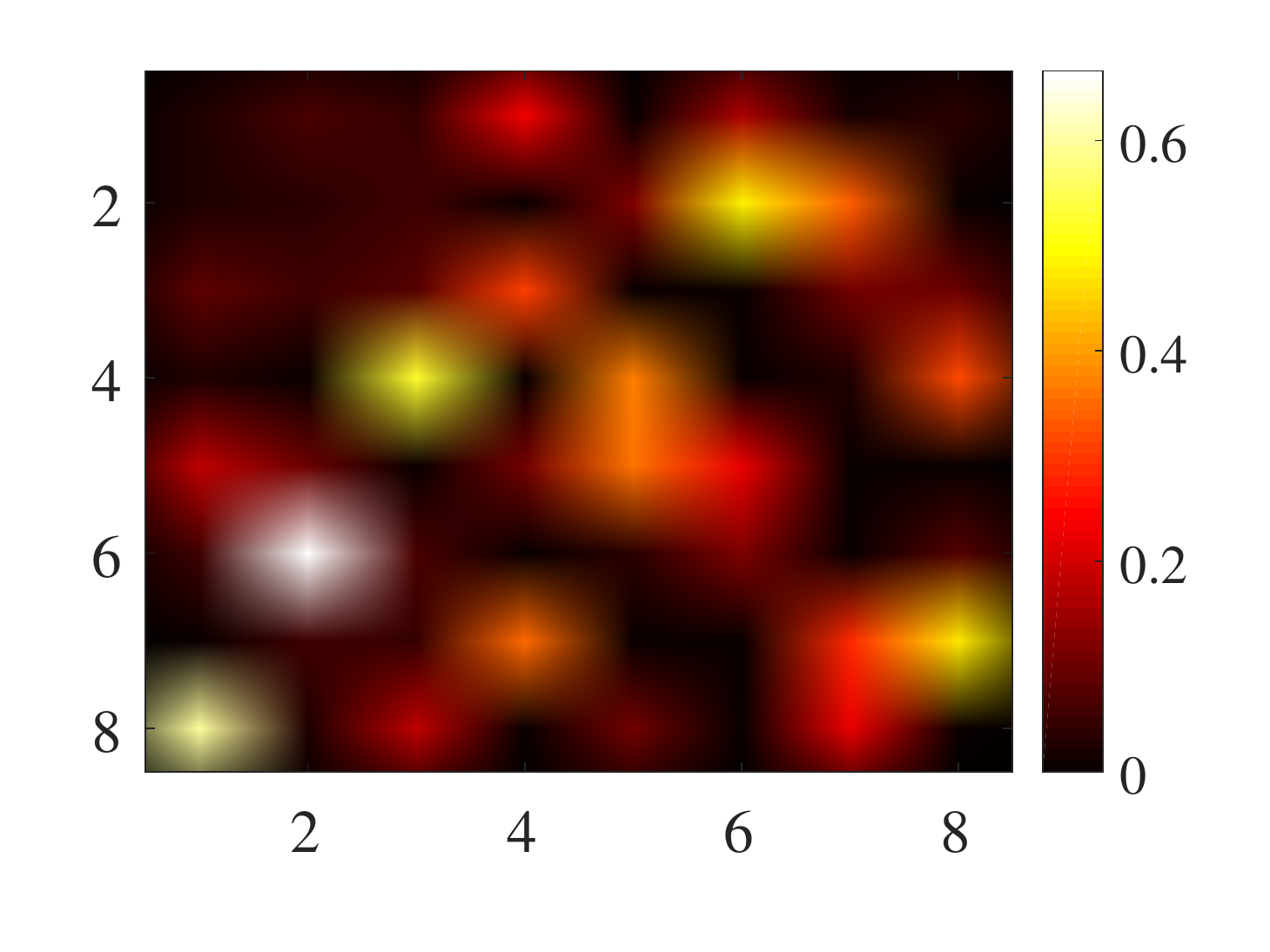}}
\end{minipage}
\begin{minipage}{0.48\linewidth}
    \centering
    \subfigure[\scriptsize{${{\bf{U}}_{\rm{T}}}$ (${\gamma} = 0.5$)}]
    {\includegraphics[width=1.0\textwidth, trim=30 30 20 20, clip]{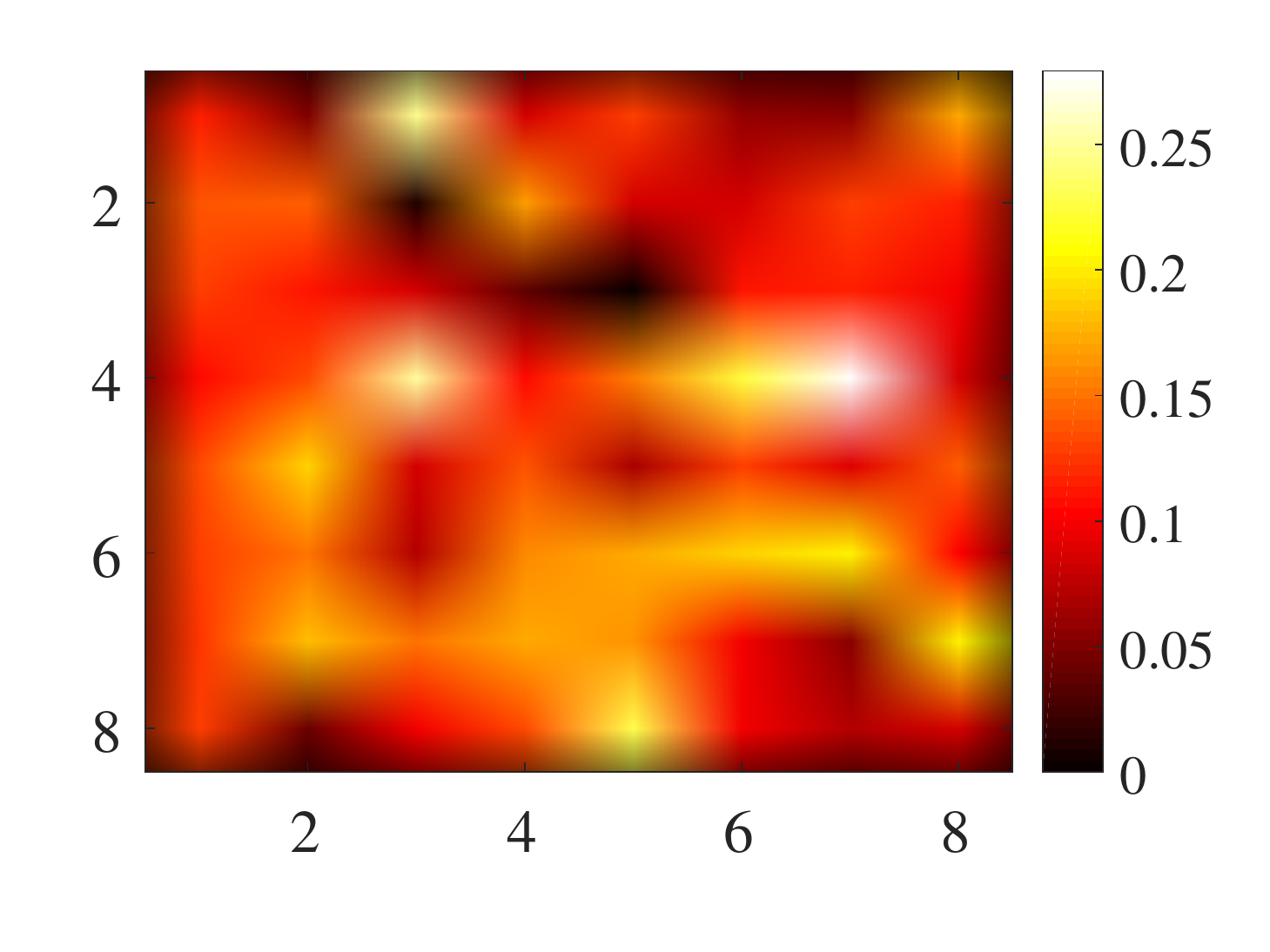}}
\end{minipage}
\begin{minipage}{0.48\linewidth}
    \centering
    \subfigure[\scriptsize{${{\bf{U}}_{\rm{A}}}$ (${\gamma} = 0$)}]
    {\includegraphics[width=1.0\textwidth, trim=30 30 20 20, clip]{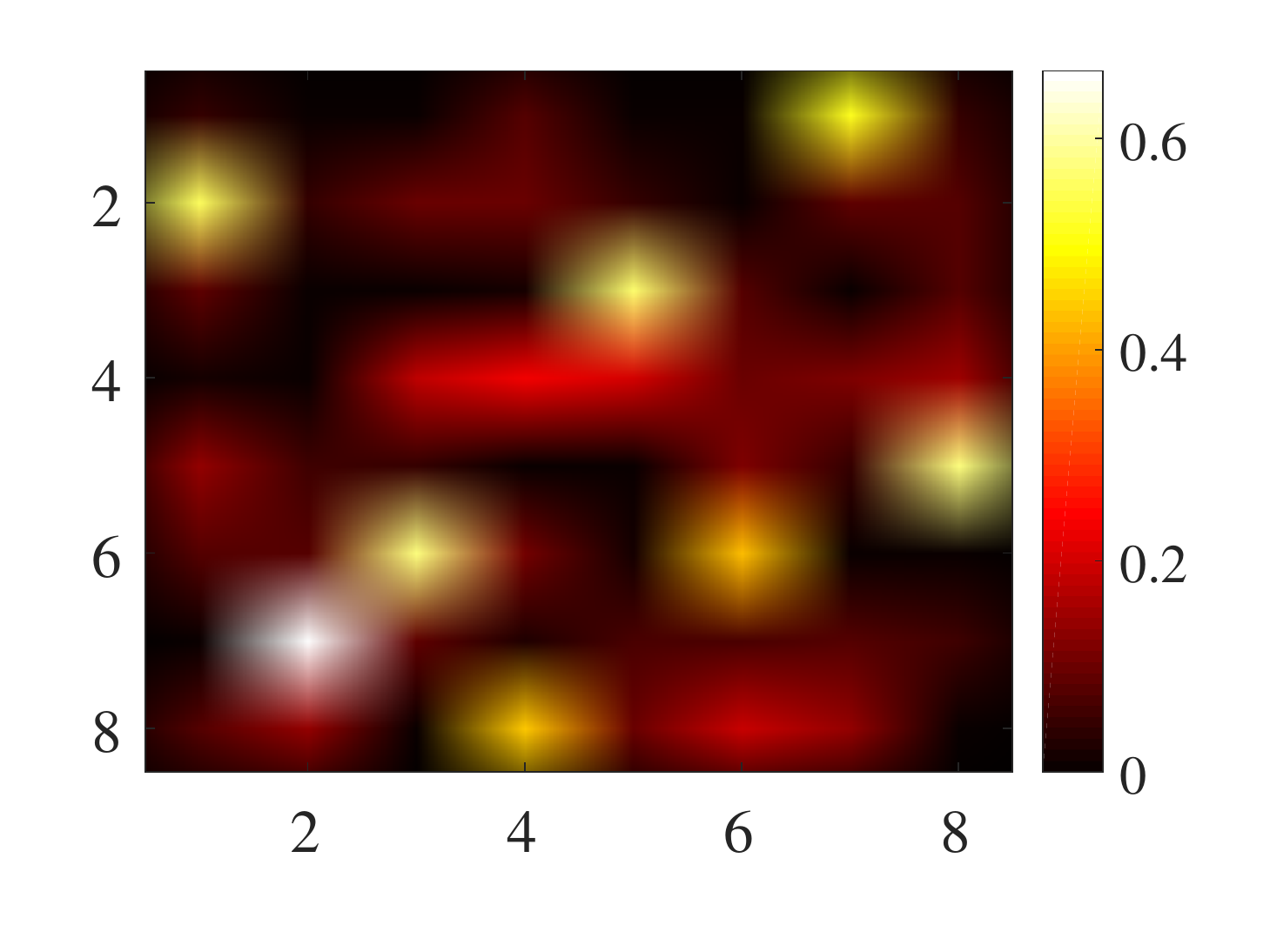}}
\end{minipage}
\begin{minipage}{0.48\linewidth}
    \centering
    \subfigure[\scriptsize{${{\bf{U}}_{\rm{A}}}$ (${\gamma} = 0.5$)}]
    {\includegraphics[width=1.0\textwidth, trim=30 30 20 20, clip]{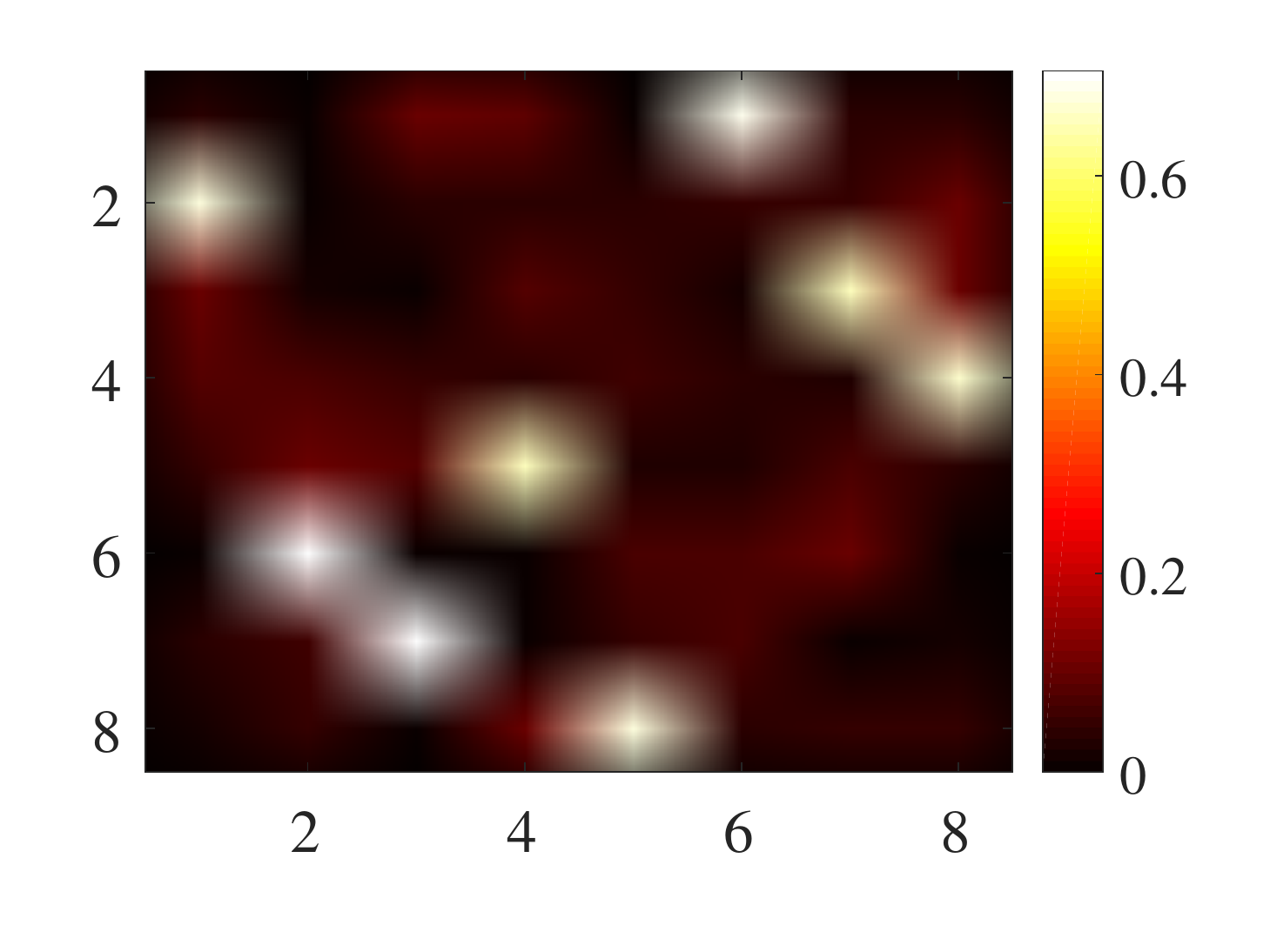}}
\end{minipage}
\caption{Example transition matrices in Case (2) with ${\gamma _{{\rm{inc}}}} \in \{ 0,0.5\}$}
\label{Trans-Vis}
\vspace{-0.5cm}
\end{figure}

In addition, we also analyzed the transition relation with different settings of $\{ {\gamma _{{\rm{inc}}}},{z_{{\rm{out}}}},{h_{{\rm{out}}}}\}$. As an example, we visualize 4 example transition matrices $\{ {{\bf{U}}_{\rm{T}}},{{\bf{U}}_{\rm{A}}}\}$ in Case (2) w.r.t. ${\gamma _{inc}} \in \{ 0.0, 0.5\}$ (${\delta_{\rm{T}}}={\delta_{\rm{A}}}=1$) in Fig.~\ref{Trans-Vis}. According to Fig.~\ref{Trans-Vis}, as ${\gamma _{inc}}$ increases, the transition relation ${{\bf{U}}_{\rm{T}}}$/${{\bf{U}}_{\rm{A}}}$ becomes more indistinct/explicit, corresponding to the decrease/increase of ${\rho _{\rm{T}}}(0)$/${\rho _{\rm{A}}}(0)$ in Fig.~\ref{Syn-Exp} (f), which validates our assumption for the \textit{inconsistency indicator} in Section~\ref{Meth}.

Please note that the synthetic graph analysis is widely adopted in the network analysis research community \cite{Girvan2002Community,He2017Joint,Qin2018Adaptive,qin2021dual}, which has several advantages for the robustness analysis beyond the direct evaluation on the real graphs. In particular, by gradually adjusting some parameters (e.g., $\{ {\gamma _{{\rm{inc}}}},{z_{{\rm{out}}}},{h_{{\rm{out}}}}\}$), the synthetic graph makes it possible to measure a model's robustness to resist the \textit{inconsistency effect} with different degrees under a quantitatively controllable condition. However, the \textit{inconsistency degree} among multiple information sources in a real graph is fixed and hard to vary. Moreover, to directly determine the type of the \textit{inconsistency} in a real graph (e.g., Case (1) to (4) in our experiments) is also challenging. In order to achieve comprehensive evaluation results, we adopted the aforementioned synthetic graph analysis as a preliminary verification of AHGR. We also applied it to a series of real graphs for further evaluation.

\subsection{Evaluation on Real Graphs}
We applied AHGR to 11 real public attributed graphs and adopted node clustering (a.k.a. community detection) and node classification as downstream tasks.

\textbf{Datasets.} Detailed statistics of the real graph datasets are shown in Table~\ref{Data}, with $N$, $E$, $M$ and $C$ as the number of nodes, edges, attributes and clusters/classes.

\begin{table}[tb]\scriptsize
\caption{Detailed statistics of the real graph datasets}
\label{Data}
\vspace{-0.3cm}
\begin{center}
\begin{tabular}{p{1.4cm}|p{0.1cm}p{0.25cm}p{0.25cm}p{0.06cm}|p{1.45cm}|p{0.25cm}p{0.25cm}p{0.25cm}p{0.1cm}}
\hline
\textbf{Datasets} & \textit{\textbf{N}} & \textit{\textbf{E}} & \textit{\textbf{M}} & \textit{\textbf{C}} & \textbf{Datasets} & \textit{\textbf{N}} & \textit{\textbf{E}} & \textit{\textbf{M}} & \textit{\textbf{C}} \\ \hline
\textit{Cornell}\tiny{\textbf{(CO)}} & 195 & 301 & 1,703 & 5 & \textit{Cora} & 2,708 & 5,429 & 1,433 & 7 \\
\textit{Texas}\tiny{\textbf{(TE)}} & 187 & 310 & 1,703 & 5 & \textit{Citeseer}\tiny{\textbf{(Cite)}} & 3,312 & 4,715 & 3,703 & 6 \\
\textit{Washington}\tiny{\textbf{(WA)}} & 230 & 395 & 1,703 & 5 & \textit{UAI2010}\tiny{\textbf{(UAI)}} & 3,067 & {\tiny{28,308}} & 4,973 & 19 \\
\textit{Wisconsin}\tiny{\textbf{(WI)}} & 265 & 510 & 1,703 & 5 & \textit{BlogCatalog}\tiny{\textbf{(BL)}} & 5,196 & {\tiny{171,743}} & 8,189 & 6 \\
\textit{Twitter}\tiny{\textbf{(TW)}} & 155 & 3,442 & 1,492 & 7 & \textit{Flickr}\tiny{\textbf{(FL)}} & 7,575 & {\tiny{239,738}} & {\tiny{12,047}} & 9 \\
\textit{Gplus}\tiny{\textbf{(GP)}} & 700 & {\tiny{28,055}} & 943 & 4 &  &  &  &  &  \\ \hline
\end{tabular}
\end{center}
\vspace{-0.5cm}
\end{table}

\textit{Cornell} (CO), \textit{Texas} (TE), \textit{Washington} (WA) and \textit{Wisconsin} (WI) are 4 sub-nets of the WebKB dataset\footnote{\scriptsize{http://www.cs.cmu.edu/afs/cs/project/theo-20/www/data/}} collected from the webpages of 4 universities, including the hyperlink relations and text content of webpages. \textit{Twitter} (TW) and \textit{Gplus} (GP) are 2 attributed ego-nets with friendship relations and user profiles, which are the subsets of the Twitter\footnote{\scriptsize{http://snap.stanford.edu/data/ego-Twitter.html}} and Google+\footnote{\scriptsize{http://snap.stanford.edu/data/ego-Gplus.html}} datasets from SNAP\footnote{\scriptsize{http://snap.stanford.edu/}}. Moreover, \textit{Cora}\footnote{\scriptsize{http://www.cs.umd.edu/{\textasciitilde}sen/lbc-proj/data/cora.tgz}} \cite{Sen2008collective} and \textit{Citeseer}\footnote{\scriptsize{http://www.cs.umd.edu/{\textasciitilde}sen/lbc-proj/data/citeseer.tgz}} (Cite) \cite{Sen2008collective} are 2 citation networks of computer science publications with different sub-fields that contains citation relationships and paper content, while \textit{UAI2010} (UAI) \cite{Sen2008collective} is a Wikipedia article reference graph including reference relations and featured lists. \textit{BlogCatalog}\footnote{\scriptsize{http://github.com/xhuang31/AANE\_MATLAB/blob/master/BlogCatalog.mat}} (BL) \cite{Huang2017Accelerated} is a subset from the blogger community BlogCatalog\footnote{\scriptsize{http://www.blogcatalog.com}} containing users' interactive relations and interest tags. \textit{Flickr}\footnote{\scriptsize{http://github.com/xhuang31/AANE\_MATLAB/blob/master/Flickr.mat}} (FL) \cite{Huang2017Accelerated} is a social network collected from the online photo sharing platform Flickr\footnote{\scriptsize{https://www.flickr.com}} with friend relations among users and photo tags of each node.

\textbf{Baselines.} We used 11 network embedding methods as baselines. DeepWalk (DW) \cite{Perozzi2014DeepWalk}, node2vec (N2V) \cite{Aditya2016node2vec}, LINE \cite{Tang2015LINE}, SDNE \cite{Wang2016Structural} GraRep \cite{Cao2015GraRep} and AROPE \cite{Zhang2018Arbitrary} are methods that exploring high-order proximities, while DNR \cite{Liang2016Modularity} and M-NMF (MNMF) \cite{Wang2017Community} are approaches integrating community structures. TADW \cite{Cheng2015Network}, AANE \cite{Huang2017Accelerated} and FSCNMF (FNMF) \cite{Bandyopadhyay2018FSCNMF} are baselines combining graph topology and attributes.

For AHGR, high-order proximities, community structures, and node attributes were adopted as available information sources, with the related variables denoted by subscripts T, C, and A. We used NMF (${\lambda _{\rm{T}}} = 5$, ${\lambda _{\rm{C}}} = {\lambda _{\rm{A}}} = 1$) and LINE (with 1st-order proximity) to derive basic embeddings, forming 2 versions of AHGR denoted as AHGR{\scriptsize{(N)}} and AHGR{\scriptsize{(L)}}).

For all the methods, we uniformly set the embedding dimensionality $K=64$ and adjusted their parameters to report best quality metrics.

\textbf{Performance Evaluation.} We adopted node clustering and classification as testing applications. For node clustering, we applied $K$Means to embeddings learned by all the methods and used normalized mutual information (NMI) \cite{Cui2017A} as well as Accuracy (AC) \cite{Cui2017A} as quality metrics. For node classification, we used SVM (with $l_2$-regularization and $l_2$-loss) implemented by LibLinear\footnote{\scriptsize{https://www.csie.ntu.edu.tw/{\textasciitilde}cjlin/liblinear/}} \cite{Fan2008LIBLINEAR} as the downstream classifier. For each dataset, 10\% of the nodes were randomly selected as the training set with the rest nodes employed for testing, where Accuracy (AC) \cite{Cui2017A} and Macro F1-Score \cite{Cui2017A} were used as evaluation metrics.

Both the clustering and classification were repeated 100 times. The average results of node clustering in terms of NMI and AC are shown in Table~\ref{Eva-Clus-NMI} and Table~\ref{Eva-Clus-AC}, while the average results of node classification w.r.t. AC and F1-Score are illustrated in Table~\ref{Eva-Clas-AC} and Table~\ref{Eva-Clas-F1}. In all the evaluation results, the best and second-best metrics are in \textbf{bold} and \underline{underlined}, respectively.

\begin{table}[tb]\scriptsize
\caption{Evaluation of Node Clustering in Terms of NMI(\%)}
\label{Eva-Clus-NMI}
\vspace{-0.3cm}
\begin{center}
\begin{tabular}{p{0.8cm}|p{0.25cm}p{0.25cm}p{0.25cm}p{0.25cm}p{0.25cm}p{0.25cm}p{0.25cm}p{0.25cm}p{0.25cm}p{0.25cm}p{0.28cm}}
\hline
 & \textbf{CO} & \textbf{TE} & \textbf{WA} & \textbf{WI} & \textbf{TW} & \textbf{GP} & \textbf{Cora} & \textbf{Cite} & \textbf{UAI} & \textbf{BL} & \textbf{FL} \\ \hline
\textit{DW} & 6.79 & 5.79 & 7.22 & 7.41 & 33.98 & 32.38 & 36.91 & 14.20 & 33.38 & 19.22 & 16.64 \\
\textit{N2V} & 6.56 & 5.55 & 6.13 & 6.81 & 32.34 & 33.12 & \underline{40.67} & 21.03 & 34.31 & 20.42 & 17.27 \\
\textit{LINE} & 12.24 & 18.56 & 21.19 & 10.79 & \underline{35.18} & 34.97 & 25.08 & 10.80 & 12.12 & 4.10 & 0.65 \\
\textit{SDNE} & 13.78 & 16.85 & 24.42 & 8.98 & 28.02 & 27.73 & 10.96 & 4.04 & 11.44 & 9.62 & 3.96 \\
\textit{GraRep} & 8.82 & 11.82 & 6.20 & 9.43 & 34.58 & 39.54 & 36.51 & 12.22 & 33.83 & 22.08 & 16.39 \\
\textit{AROPE} & 9.08 & 10.29 & 9.63 & 6.36 & 29.32 & 19.45 & 8.85 & 4.54 & 13.63 & 14.37 & 8.40 \\
\textit{DNR} & 11.91 & 18.70 & 24.29 & 9.46 & 32.45 & 25.47 & 16.09 & 6.48 & 5.66 & 13.28 & 4.42 \\
\textit{MNMF} & 13.00 & 18.28 & 23.19 & 8.98 & 33.85 & \underline{41.21} & 10.46 & 5.34 & 19.52 & 17.25 & 14.76 \\
\textit{TADW} & 11.65 & 7.58 & 11.60 & 13.19 & 25.46 & 7.75 & 28.28 & \underline{21.24} & 25.19 & 7.85 & 2.27 \\
\textit{AANE} & 30.66 & 32.43 & 38.32 & \underline{42.46} & 30.33 & 37.72 & 15.29 & 18.10 & 30.98 & \underline{28.26} & \underline{39.30} \\
\textit{FNMF} & 12.52 & 16.73 & 11.78 & 12.37 & 11.13 & 25.30 & 11.56 & 17.52 & \underline{43.70} & 1.46 & 0.37 \\ \hline
AHGR\tiny(N) & \textbf{35.12} & \textbf{35.49} & \textbf{41.41} & \textbf{45.70} & \textbf{36.69} & \textbf{47.50} & \textbf{41.07} & \textbf{24.66} & \textbf{46.74} & \textbf{33.80} & \textbf{40.23} \\
AHGR\tiny{(L)} & \underline{34.94} & \underline{33.58} & \underline{39.15} & 41.61 & 34.20 & 35.76 & 33.36 & 17.79 & 37.63 & 21.05 & 17.32 \\ \hline
\end{tabular}
\end{center}
\vspace{-0.5cm}
\end{table}

\begin{table}[tb]\scriptsize
\caption{Evaluation of Node Clustering in Terms of AC(\%)}
\label{Eva-Clus-AC}
\vspace{-0.3cm}
\begin{center}
\begin{tabular}{p{0.8cm}|p{0.25cm}p{0.25cm}p{0.25cm}p{0.25cm}p{0.25cm}p{0.25cm}p{0.25cm}p{0.25cm}p{0.25cm}p{0.25cm}p{0.28cm}}
\hline
 & \textbf{CO} & \textbf{TE} & \textbf{WA} & \textbf{WI} & \textbf{TW} & \textbf{GP} & \textbf{Cora} & \textbf{Cite} & \textbf{UAI} & \textbf{BL} & \textbf{FL} \\ \hline
\textit{DW} & 38.31 & 50.37 & 44.60 & 43.65 & 41.89 & 56.96 & 51.55 & 40.79 & 36.49 & 35.44 & 30.87 \\
\textit{N2V} & 37.50 & 47.02 & 40.97 & 38.92 & 36.74 & 54.53 & \underline{54.95} & \underline{44.00} & 37.69 & 36.81 & 31.42 \\
\textit{LINE} & 38.98 & 54.79 & 56.16 & 43.74 & 42.36 & 56.58 & 42.37 & 26.96 & 16.49 & 25.24 & 13.09 \\
\textit{SDNE} & 42.12 & 54.95 & \underline{62.30} & 47.43 & 36.06 & 55.27 & 31.25 & 22.51 & 19.17 & 26.88 & 15.52 \\
\textit{GraRep} & 32.61 & 35.34 & 31.78 & 32.36 & 43.12 & 54.11 & 51.12 & 33.95 & 37.86 & 38.79 & 29.18 \\
\textit{AROPE} & 42.95 & 56.02 & 48.91 & 46.51 & 37.63 & 55.87 & 32.68 & 23.05 & 21.24 & 28.18 & 18.27 \\
\textit{DNR} & 37.59 & 52.42 & 55.34 & 42.71 & \underline{44.57} & 53.31 & 33.56 & 23.67 & 12.89 & 32.57 & 18.52 \\
\textit{MNMF} & 39.21 & 55.00 & 60.22 & 45.55 & 39.32 & 54.42 & 32.67 & 23.34 & 24.36 & 33.90 & 28.35 \\
\textit{TADW} & 47.69 & 57.30 & 50.93 & 50.57 & 42.93 & 43.10 & 44.88 & 40.35 & 27.58 & 23.26 & 14.15 \\
\textit{AANE} & \underline{50.77} & 56.25 & 54.76 & \underline{59.85} & 43.29 & \underline{65.31} & 33.90 & 40.04 & 29.73 & \underline{45.14} & \underline{38.57} \\
\textit{FNMF} & 48.14 & \underline{58.86} & 50.83 & 50.42 & 35.01 & 62.25 & 32.12 & 39.57 & 40.64 & 18.92 & 11.81 \\ \hline
AHGR\tiny{(N)} & \textbf{58.56} & \textbf{61.21} & \textbf{62.35} & \textbf{60.09} & \textbf{44.62} & 62.11 & \textbf{57.06} & \textbf{45.27} & \textbf{44.87} & \textbf{50.42} & \textbf{52.64} \\
AHGR\tiny{(L)} & 49.44 & 49.15 & 51.81 & 59.26 & 40.85 & \textbf{65.34} & 51.14 & 32.34 & \underline{41.09} & 37.33 & 31.72 \\ \hline
\end{tabular}
\end{center}
\vspace{-0.5cm}
\end{table}

\begin{table}[tb]\scriptsize
\caption{Evaluation of Node Classification in Terms of AC(\%)}
\label{Eva-Clas-AC}
\vspace{-0.3cm}
\begin{center}
\begin{tabular}{p{0.8cm}|p{0.25cm}p{0.25cm}p{0.25cm}p{0.25cm}p{0.25cm}p{0.25cm}p{0.25cm}p{0.25cm}p{0.25cm}p{0.25cm}p{0.28cm}}
\hline
 & \textbf{CO} & \textbf{TE} & \textbf{WA} & \textbf{WI} & \textbf{TW} & \textbf{GP} & \textbf{Cora} & \textbf{Cite} & \textbf{UAI} & \textbf{BL} & \textbf{FL} \\ \hline
\textit{DW} & 32.31 & 47.34 & 38.85 & 40.66 & 48.42 & 85.91 & 68.24 & 45.32 & 45.71 & 59.63 & 44.54 \\
\textit{N2V} & 31.85 & 47.40 & 39.83 & 40.60 & 48.54 & 87.10 & 72.53 & 50.18 & 48.50 & 59.94 & 45.99 \\
\textit{LINE} & 37.19 & 57.39 & 53.07 & 48.64 & 46.88 & 90.08 & 71.06 & 44.90 & 35.05 & 32.35 & 12.39 \\
\textit{SDNE} & 42.12 & 54.95 & 62.35 & 47.43 & 36.06 & 55.27 & 31.25 & 22.51 & 19.17 & 56.37 & 31.73 \\
\textit{GraRep} & 35.78 & 51.24 & 39.79 & 43.27 & \underline{51.29} & 87.79 & \underline{73.96} & 48.62 & 52.45 & 65.87 & 50.24 \\
\textit{AROPE} & 37.69 & 55.37 & 50.50 & 46.30 & 48.67 & \underline{93.90} & 65.14 & 43.13 & 45.42 & 67.12 & 57.17 \\
\textit{DNR} & 38.46 & 57.37 & 58.63 & 50.24 & 46.88 & 91.88 & 44.49 & 27.31 & 17.90 & 40.79 & 17.80 \\
\textit{MNMF} & 36.71 & 57.92 & 47.11 & 47.09 & 50.35 & 91.33 & 69.68 & 47.06 & 45.11 & 66.45 & 53.29 \\
\textit{TADW} & 43.28 & 51.51 & 50.31 & 51.11 & 46.74 & 84.12 & 66.17 & 57.06 & 55.59 & \underline{89.76} & 56.65 \\
\textit{AANE} & \underline{59.49} & 65.57 & \underline{70.42} & \underline{72.63} & 47.32 & 84.86 & 64.67 & \underline{63.43} & 64.47 & 82.19 & \textbf{85.97} \\
\textit{FNMF} & 54.51 & 64.18 & 64.55 & 64.43 & 39.04 & 82.46 & 60.65 & 62.60 & \underline{68.43} & 73.11 & 49.45 \\ \hline
AHGR\tiny{(N)} & \textbf{62.65} & \textbf{69.54} & \textbf{73.57} & \textbf{76.33} & \textbf{51.47} & \textbf{94.39} & \textbf{75.15} & \textbf{65.10} & \textbf{68.91} & \textbf{89.97} & \underline{83.01} \\
AHGR\tiny{(L)} & 58.43 & \underline{67.37} & 66.45 & 71.34 & 49.21 & 85.32 & 71.69 & 51.71 & 50.46 & 60.31 & 46.13 \\ \hline
\end{tabular}
\end{center}
\vspace{-0.5cm}
\end{table}

\begin{table}[tb]\scriptsize
\caption{Evaluation of Node Classification in Terms of F1-Score(\%)}
\label{Eva-Clas-F1}
\vspace{-0.3cm}
\begin{center}
\begin{tabular}{p{0.8cm}|p{0.25cm}p{0.25cm}p{0.25cm}p{0.25cm}p{0.25cm}p{0.25cm}p{0.25cm}p{0.25cm}p{0.25cm}p{0.25cm}p{0.28cm}}
\hline
 & \textbf{CO} & \textbf{TE} & \textbf{WA} & \textbf{WI} & \textbf{TW} & \textbf{GP} & \textbf{Cora} & \textbf{Cite} & \textbf{UAI} & \textbf{BL} & \textbf{FL} \\ \hline
\textit{DW} & 20.05 & 21.15 & 20.79 & 24.22 & 25.87 & 53.69 & 66.86 & 41.95 & 37.68 & 59.11 & 43.61 \\
\textit{N2V} & 19.55 & 20.32 & 21.03 & 24.61 & 18.29 & 51.44 & 70.82 & 45.96 & 39.42 & 59.30 & 44.71 \\
\textit{LINE} & 24.10 & 27.36 & 28.17 & 28.72 & 25.18 & 55.85 & 69.20 & 40.51 & 24.33 & 27.89 & 10.42 \\
\textit{SDNE} & 22.63 & 25.69 & 30.08 & 26.02 & 24.56 & 64.81 & 22.41 & 10.32 & 20.18 & 55.63 & 27.59 \\
\textit{GraRep} & 25.20 & 29.48 & 24.14 & 27.43 & 27.68 & 58.13 & \underline{72.67} & 45.02 & 42.96 & 65.22 & 49.73 \\
\textit{AROPE} & 24.44 & 28.89 & 26.89 & 28.33 & 25.74 & \textbf{71.52} & 63.47 & 39.36 & 35.09 & 65.79 & 55.58 \\
\textit{DNR} & 19.18 & 25.70 & 28.32 & 22.59 & 24.57 & 63.09 & 29.26 & 17.28 & 7.71 & 39.61 & 13.40 \\
\textit{MNMF} & 22.16 & 30.80 & 24.57 & 28.87 & \underline{27.10} & 66.16 & 67.16 & 42.78 & 35.81 & 65.54 & 52.46 \\
\textit{TADW} & 27.75 & 21.81 & 27.58 & 30.11 & 26.60 & 50.03 & 63.83 & 52.97 & 43.82 & \underline{89.59} & 55.90 \\
\textit{AANE} & \underline{38.50} & 31.55 & \underline{38.40} & 44.08 & 20.17 & 43.91 & 60.30 & 55.31 & 45.86 & 81.75 & \textbf{85.63} \\
\textit{FNMF} & 35.02 & \underline{34.66} & 37.39 & 40.55 & 19.68 & 42.73 & 56.99 & \underline{55.97} & \underline{53.43} & 72.61 & 46.16 \\ \hline
AHGR\tiny{(N)} & \textbf{42.13} & \textbf{38.11} & \textbf{42.80} & \textbf{48.59} & \textbf{27.71} & \underline{68.84} & \textbf{72.76} & \textbf{59.35} & \textbf{53.80} & \textbf{89.82} & \underline{82.65} \\
AHGR\tiny{(L)} & 36.79 & 33.79 & 35.61 & \underline{44.26} & 20.99 & 50.58 & 69.90 & 45.79 & 36.93 & 59.64 & 44.89 \\ \hline
\end{tabular}
\end{center}
\vspace{-0.5cm}
\end{table}

For node clustering, AHGR{\scriptsize{(N)}} achieves the best performance on all the datasets in terms of NMI, while it performs the best on 10 of the 11 datasets in terms of AC (with average improvements of 9.57\% and 7.92\% for NMI and AC w.r.t. the second-best baseline). For node classification, AHGR{\scriptsize{(N)}} performs the best on 10 of the 11 datasets and the second-best on the rest dataset in terms of AC, while it has the best performance on 9 of the 11 datasets and the second-best performance on the rest 2 datasets in terms of F1-Score (with average improvement of 2.70\% and 5.60\% for AC and F1-Score w.r.t. the second-best method). Furthermore, AHGR{\tiny{(L)}} also achieves the performances competitive to other baselines for both node clustering and classification.

There remains a gap between the performance of AHGR{\scriptsize{(N)}} and AHGR{\scriptsize{(L)}}, indicating that the selection of the basic embedding method may affect the final result. In fact, NMF and LINE (w.r.t. AHGR{\scriptsize{(N)}} and AHGR{\scriptsize{(L)}}) are two distinct types of embedding approaches. On the one hand, NMF learns low-dimensional representations based on the entire topology of each reweighted graph (i.e., factorization of the entire adjacency matrix), but with relatively high complexity. On the other hand, LINE is a sampling-based approach with much lower complexity, which learns the embedding only based on the sampled edges and is more appropriate for sparse graphs. However, not all auxiliary weighted graphs are sparse enough for the sampling-based approaches to achieve relatively high performance, in which the sampling procedure is more likely to lose some information regarding the reweighted topology. Hence, there exists a compromise between quality and efficiency for the selection of the basic embedding method.

\textbf{Parameter Analysis.} In our experiments, we adjusted parameters for both AHGR{\scriptsize{(N)}} and AHGR{\scriptsize{(L)}} by setting $h \in \{ 1, \cdots ,8\}$ and ${\delta _{\rm{T}}},{\delta _{\rm{C}}},{\delta _{\rm{A}}},\delta \in \{ 1,5,10\}$. Due to space limit, we omit details of the parameter adjustment and directly list the recommended parameter settings (w.r.t. Tables~\ref{Eva-Clus-NMI}, \ref{Eva-Clus-AC}, \ref{Eva-Clas-AC} and \ref{Eva-Clas-F1}) in Table~\ref{param}, where ${\delta_*}$ represents the settings of $\{{\delta _{\rm{T}}},{\delta _{\rm{C}}},{\delta _{\rm{A}}}\}$ in sequence. 

\begin{table}[]\tiny
\caption{Parameter setting of AHGR in real graph evaluation}
\label{param}
\vspace{-0.3cm}
\begin{center}
\begin{tabular}{p{0.65cm}|p{0.1cm}|p{0.2cm}p{0.2cm}p{0.2cm}p{0.2cm}p{0.2cm}p{0.2cm}p{0.2cm}p{0.2cm}p{0.2cm}p{0.2cm}p{0.3cm}}
\hline
 &  & \textbf{CO} & \textbf{TE} & \textbf{WA} & \textbf{WI} & \textbf{TW} & \textbf{GP} & \textbf{Cora} & \textbf{Cite} & \textbf{UAI} & \textbf{BL} & \textbf{FL} \\ \hline
\multirow{3}{*}{\textbf{AHGR}(N)} & $h$ & 1 & 1 & 1 & 1 & 3 & 2 & 4 & 1 & 2 & 2 & 2 \\
 & ${{\delta}_*}$ & 5,5,1 & 10,10,1 & 5,5,1 & 10,10,1 & 10,1,10 & 1,1,1 & 1,1,1 & 5,5,1 & 10,1,1 & 5,1,1 & 10,1,10 \\
 & $\delta$ & 1 & 1 & 1 & 1 & 10 & 10 & 10 & 10 & 10 & 1 & 10 \\ \hline
\multirow{3}{*}{\textbf{AHGR}(L)} & $h$ & 1 & 1 & 1 & 1 & 2 & 2 & 2 & 8 & 2 & 3 & 2 \\
 & ${{\delta}_*}$ & 10,10,1 & 10,10,1 & 10,10,1 & 10,10,1 & 5,1,10 & 10,5,1 & 1,1,1 & 5,5,1 & 1,1,1 & 1,1,1 & 10,10,1 \\
 & $\delta$ & 10 & 10 & 10 & 10 & 10 & 10 & 10 & 10 & 1 & 10 & 10 \\ \hline
\end{tabular}
\end{center}
\vspace{-0.5cm}
\end{table}

\begin{figure*}[tb]
\centering
\begin{minipage}{0.24\linewidth}
    \centering
    \subfigure[AHGR{\tiny{(N)}} on \textit{WI}]
    {\includegraphics[width=1.0\textwidth, trim=20 0 45 5, clip]{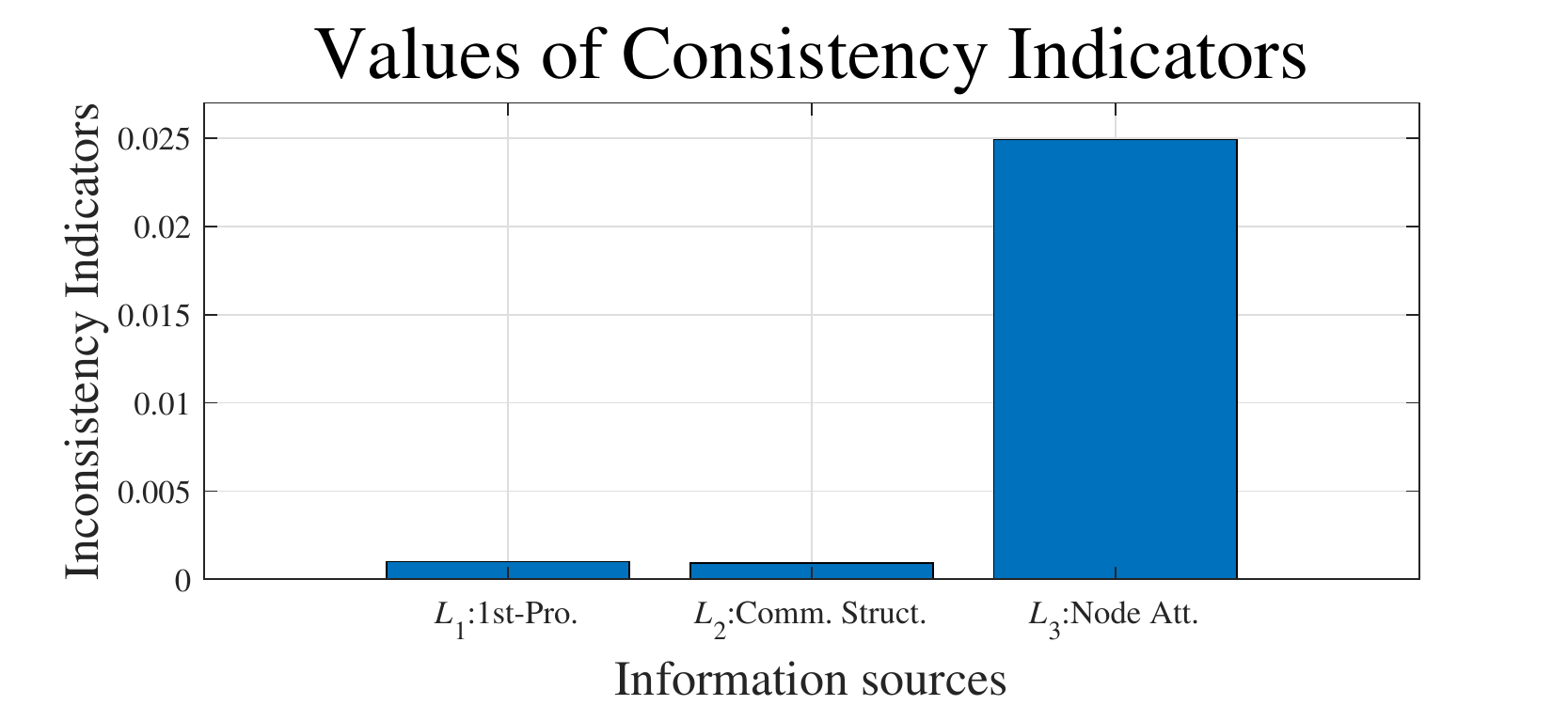}}
\end{minipage}
\begin{minipage}{0.24\linewidth}
    \centering
    \subfigure[AHGR{\tiny{(N)}} on \textit{WI}]
    {\includegraphics[width=1.0\textwidth, trim=20 0 45 5, clip]{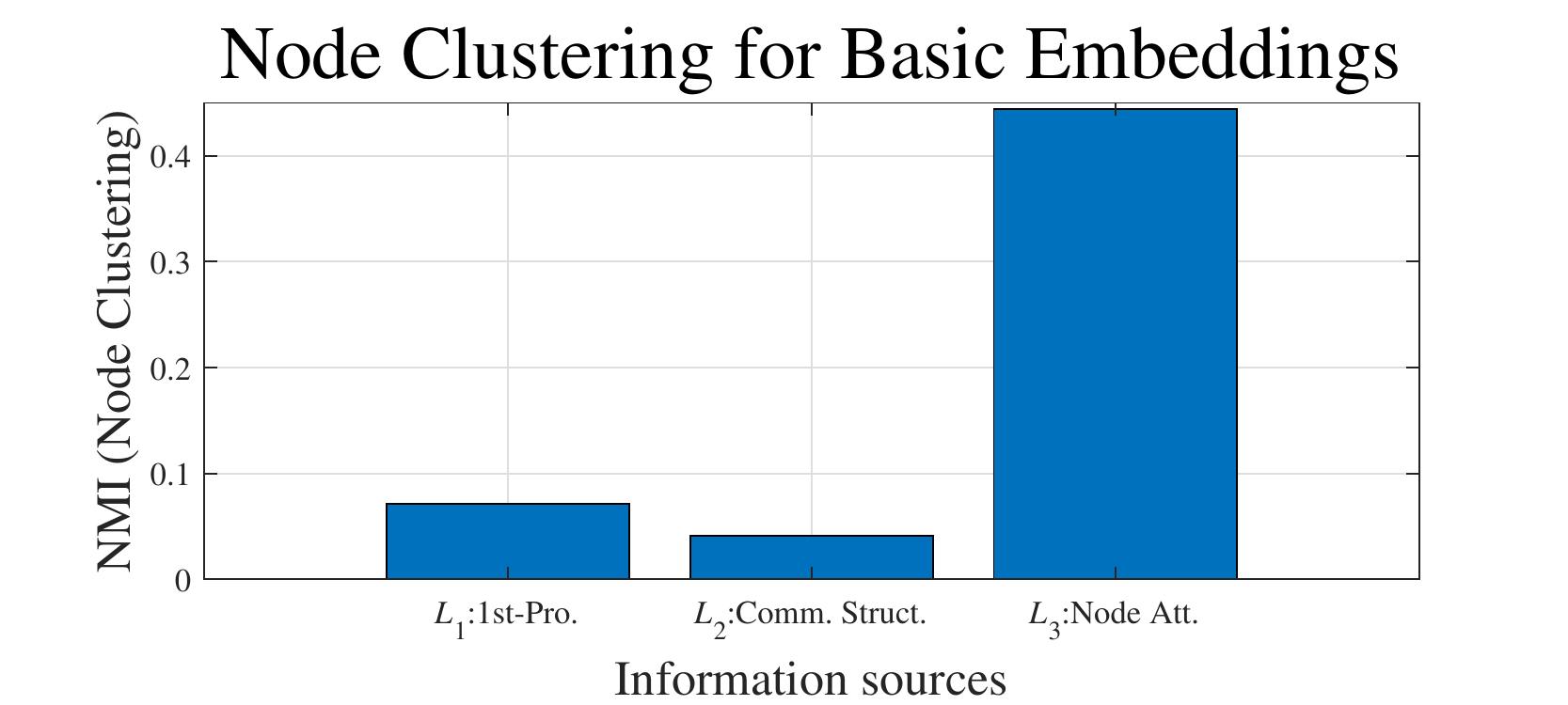}}
\end{minipage}
\begin{minipage}{0.24\linewidth}
    \centering
    \subfigure[AHGR{\tiny{(N)}} on \textit{WI}]
    {\includegraphics[width=1.0\textwidth, trim=20 0 45 5, clip]{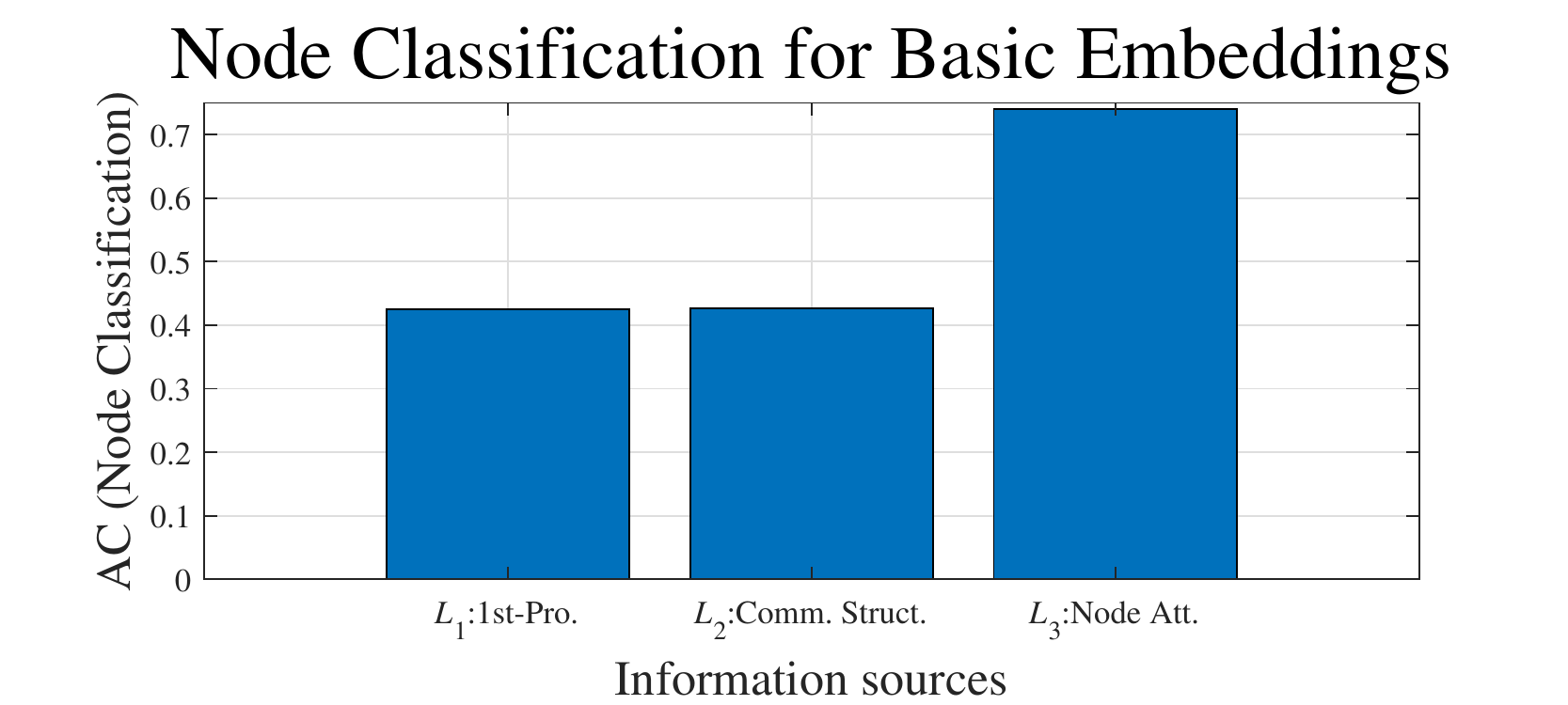}}
\end{minipage}
\begin{minipage}{0.24\linewidth}
    \centering
    \subfigure[AHGR{\tiny{(L)}} on \textit{WI}]
    {\includegraphics[width=1.0\textwidth, trim=20 0 45 5, clip]{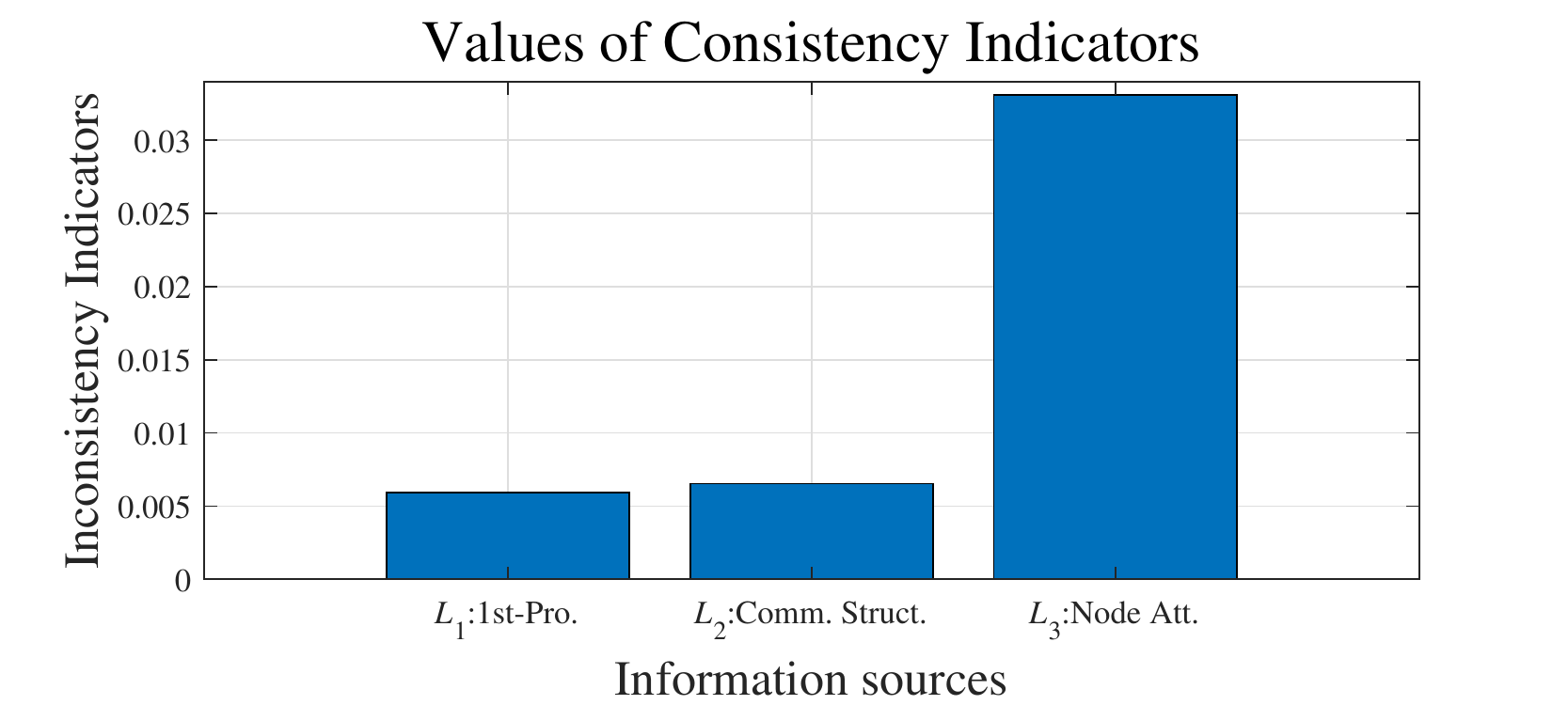}}
\end{minipage}
\begin{minipage}{0.24\linewidth}
    \centering
    \subfigure[AHGR{\tiny{(L)}} on \textit{WI}]
    {\includegraphics[width=1.0\textwidth, trim=20 0 45 5, clip]{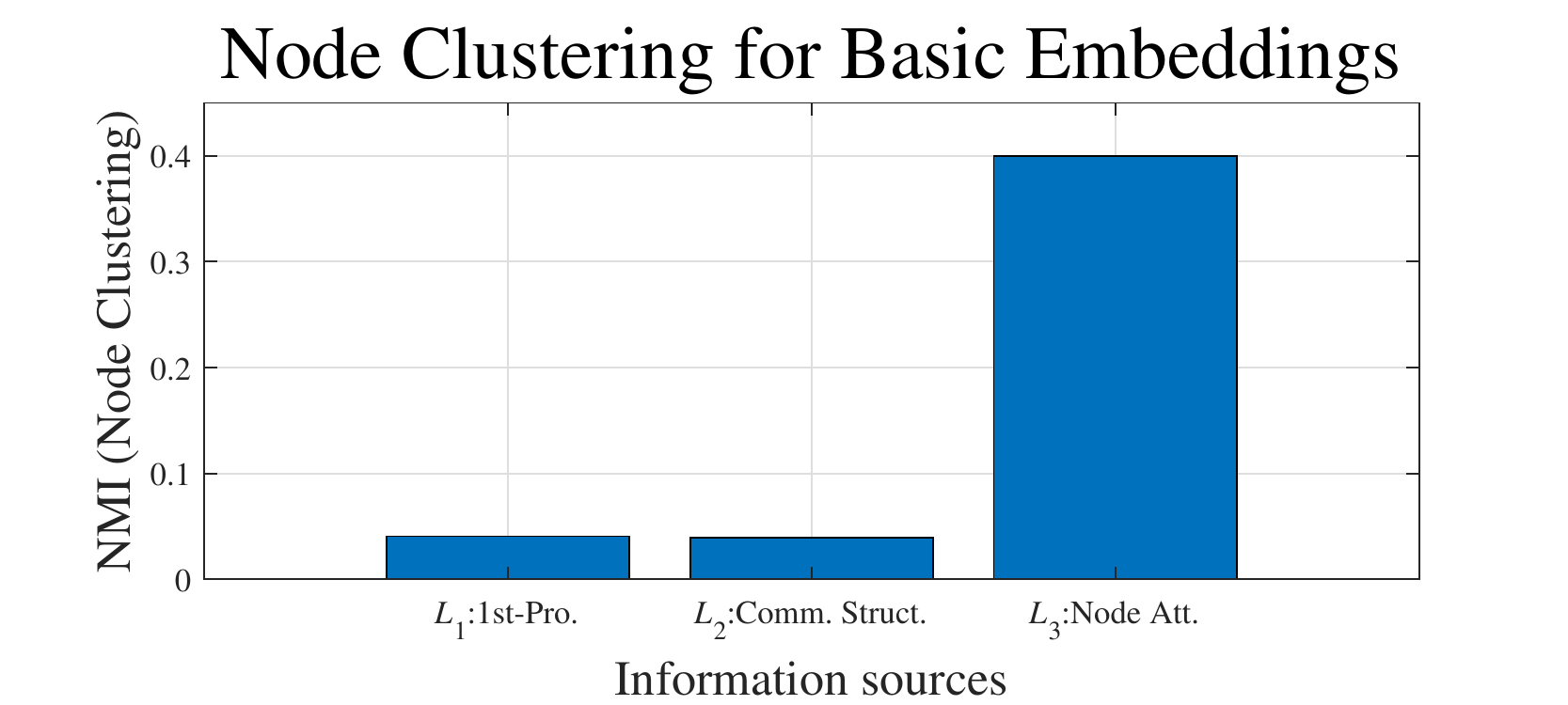}}
\end{minipage}
\begin{minipage}{0.24\linewidth}
    \centering
    \subfigure[AHGR{\tiny{(L)}} on \textit{WI}]
    {\includegraphics[width=1.0\textwidth, trim=20 0 45 5, clip]{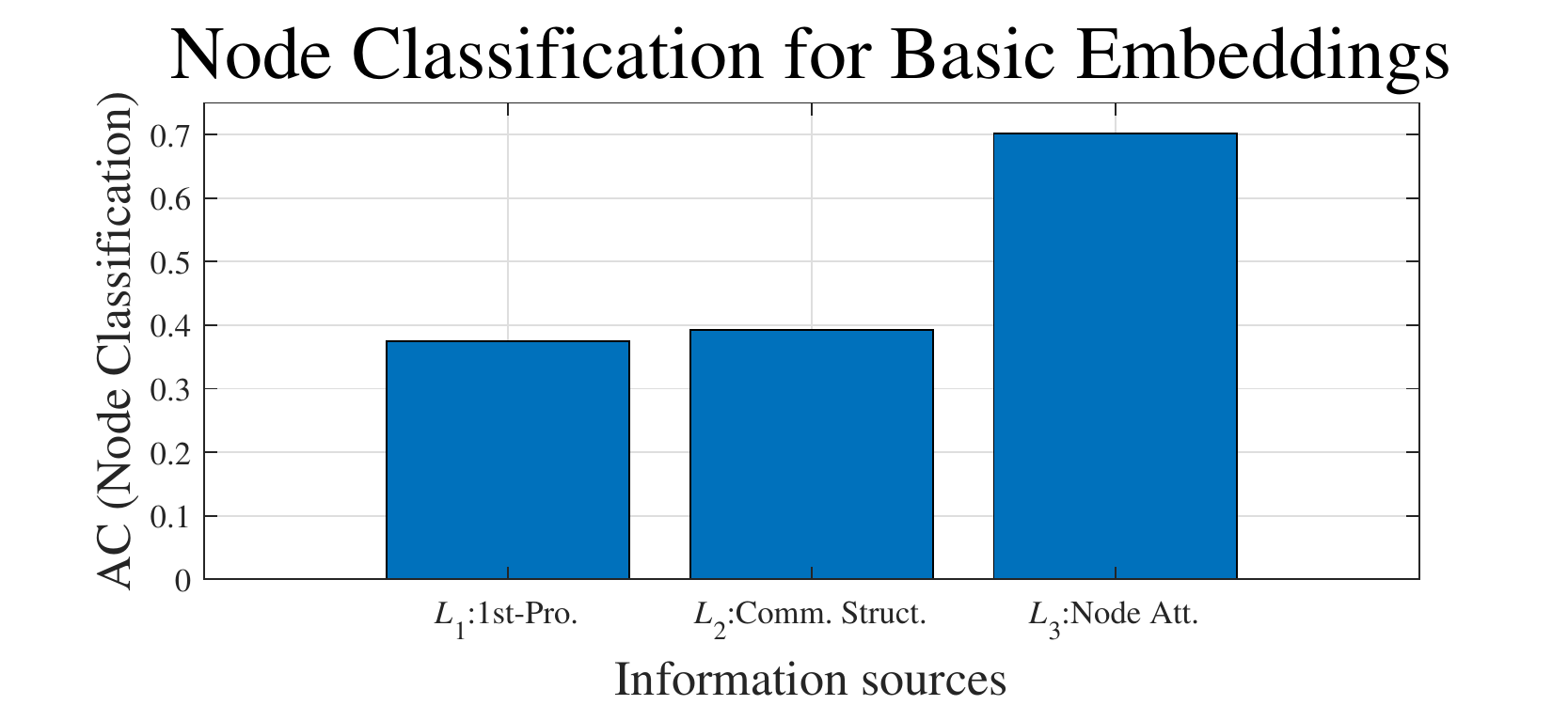}}
\end{minipage}
\begin{minipage}{0.24\linewidth}
    \centering
    \subfigure[AHGR{\tiny{(N)}} on \textit{TW}]
    {\includegraphics[width=1.0\textwidth, trim=20 0 45 5, clip]{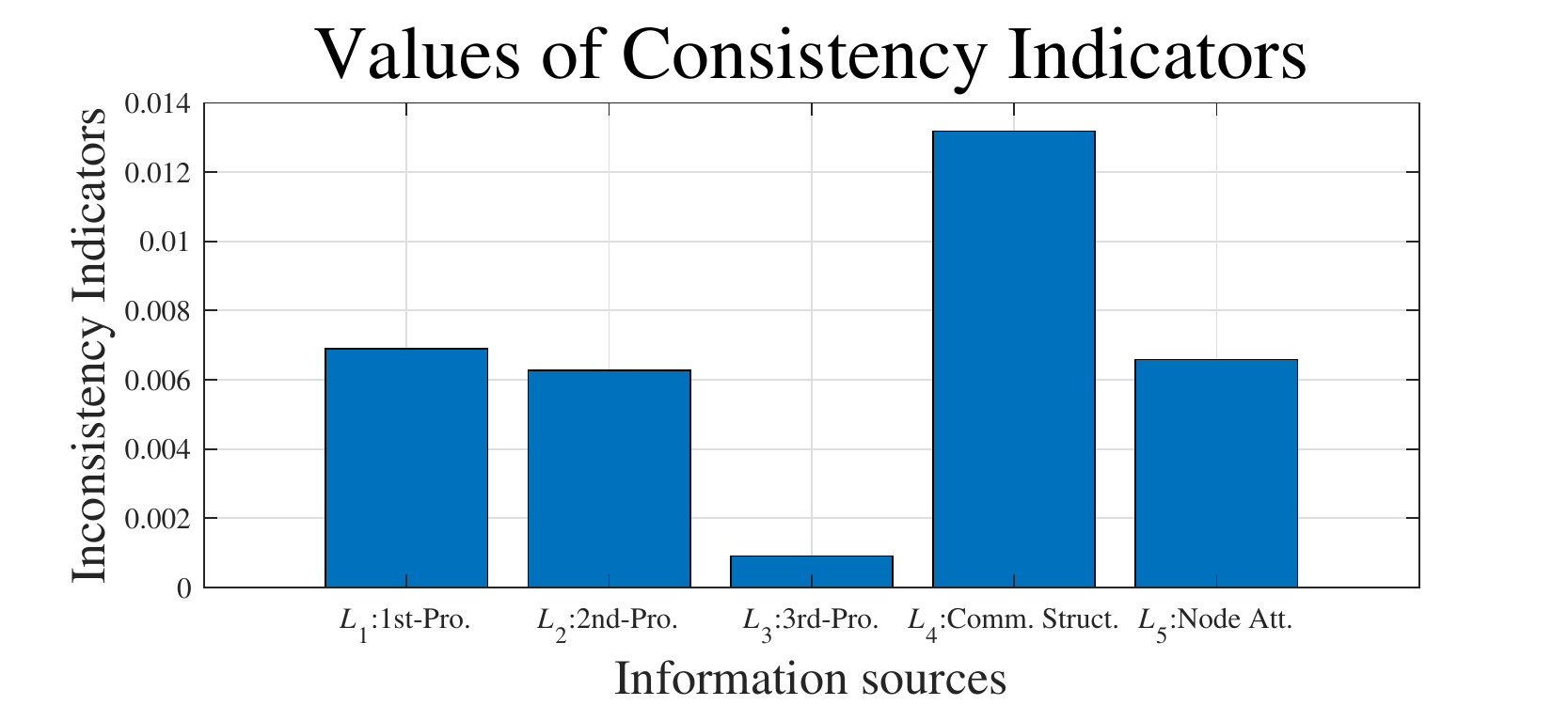}}
\end{minipage}
\begin{minipage}{0.24\linewidth}
    \centering
    \subfigure[AHGR{\tiny{(N)}} on \textit{TW}]
    {\includegraphics[width=1.0\textwidth, trim=20 0 45 5, clip]{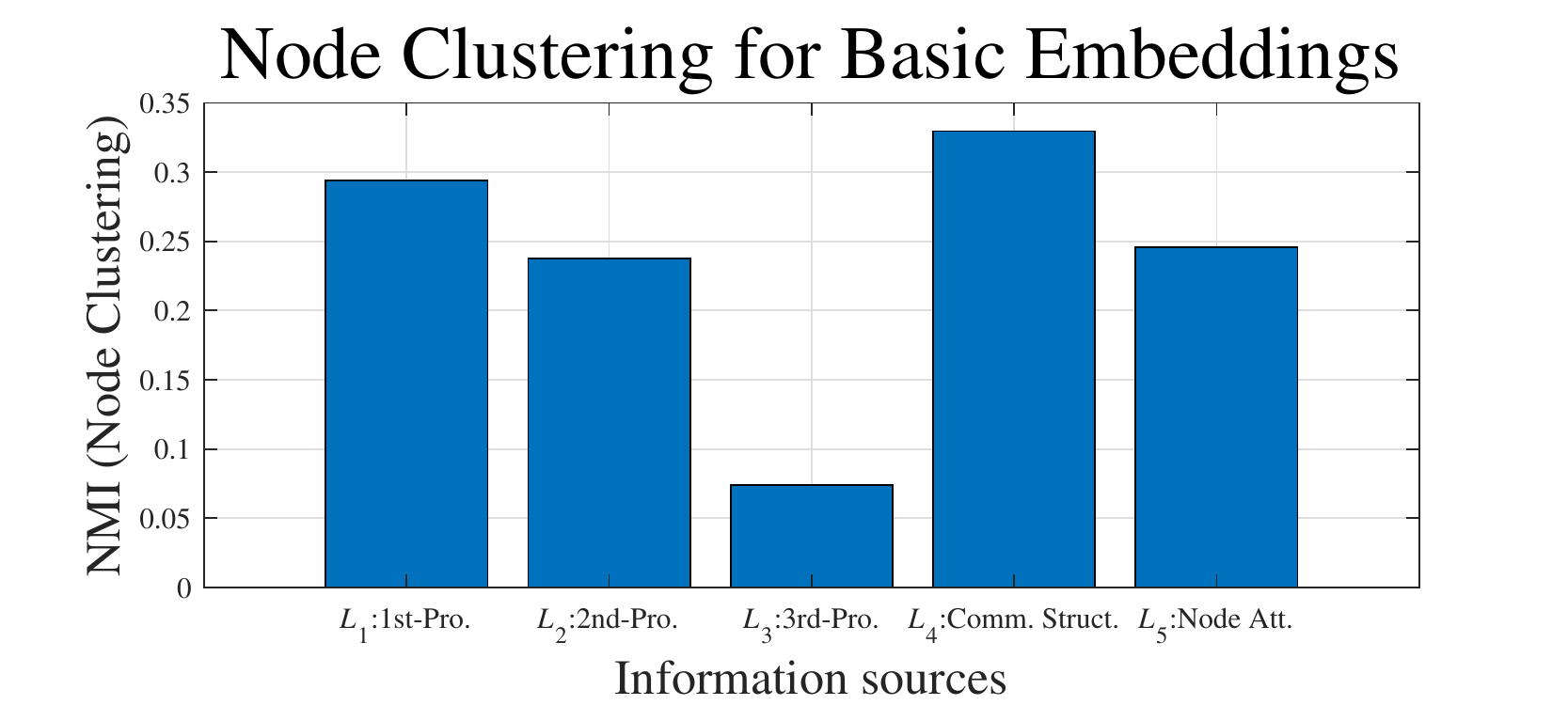}}
\end{minipage}
\begin{minipage}{0.24\linewidth}
    \centering
    \subfigure[AHG{R\tiny{(N)}}, \textit{TW}]
    {\includegraphics[width=1.0\textwidth, trim=20 0 45 5, clip]{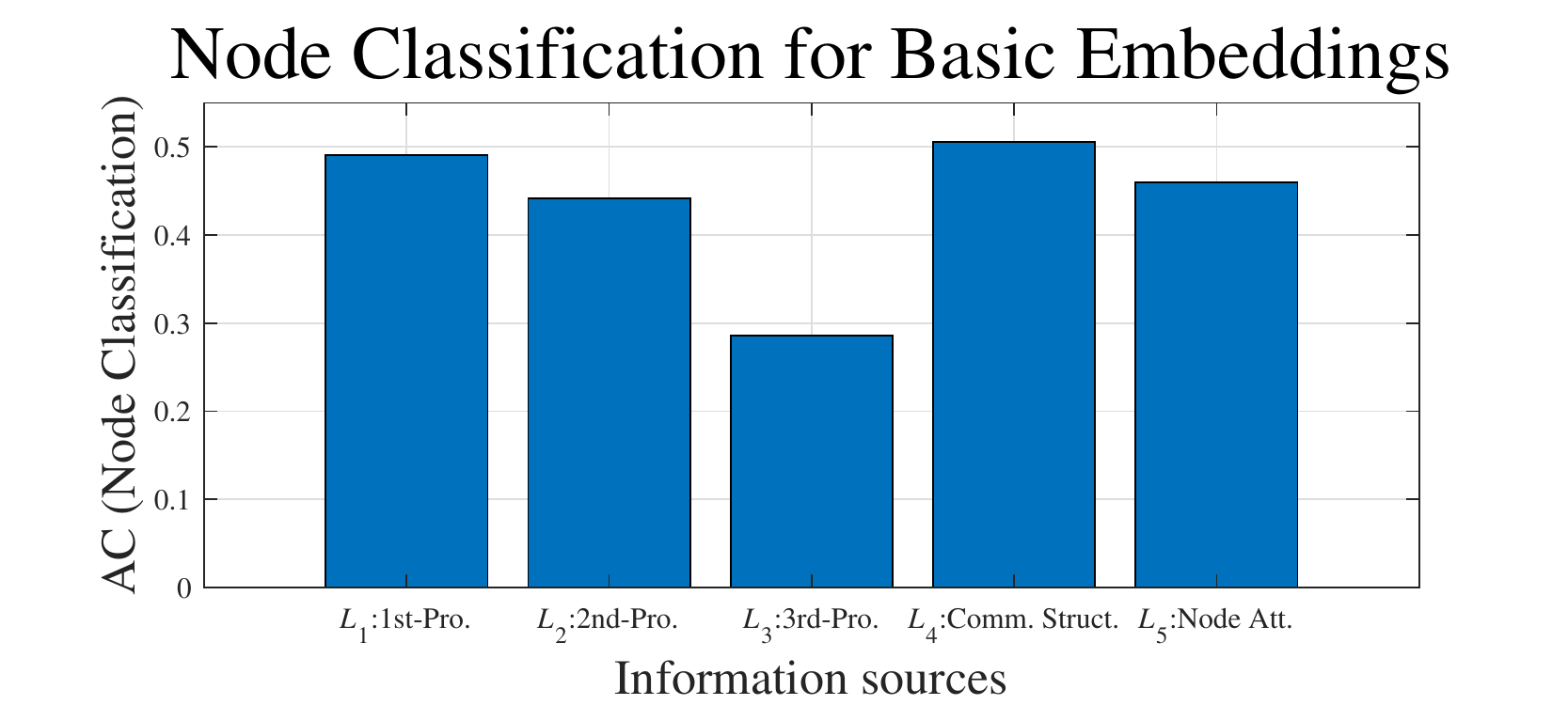}}
\end{minipage}
\begin{minipage}{0.24\linewidth}
    \centering
    \subfigure[AHGR{\tiny{(L)}} on \textit{TW}]
    {\includegraphics[width=1.0\textwidth, trim=20 0 45 5, clip]{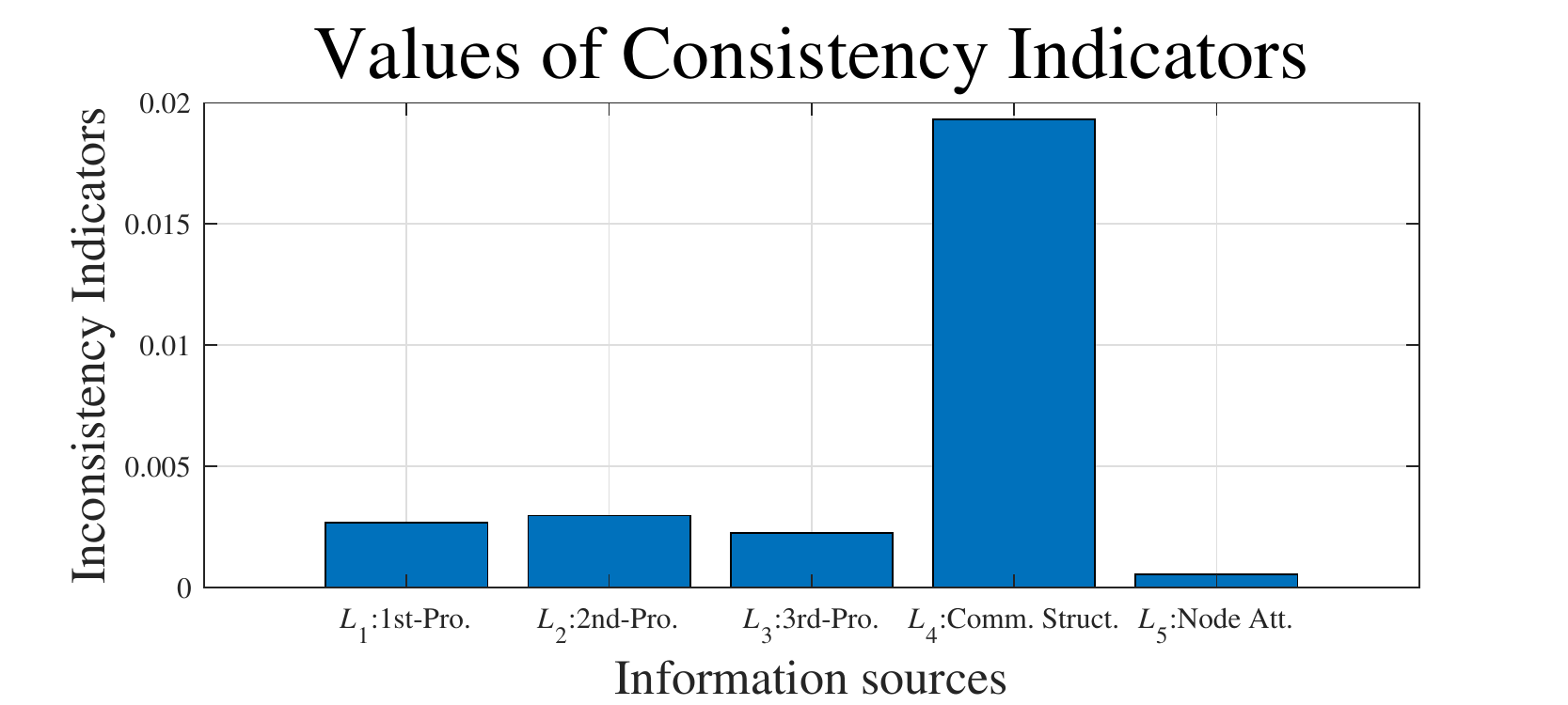}}
\end{minipage}
\begin{minipage}{0.24\linewidth}
    \centering
    \subfigure[AHGR{\tiny{(L)}} on \textit{TW}]
    {\includegraphics[width=1.0\textwidth, trim=20 0 45 5, clip]{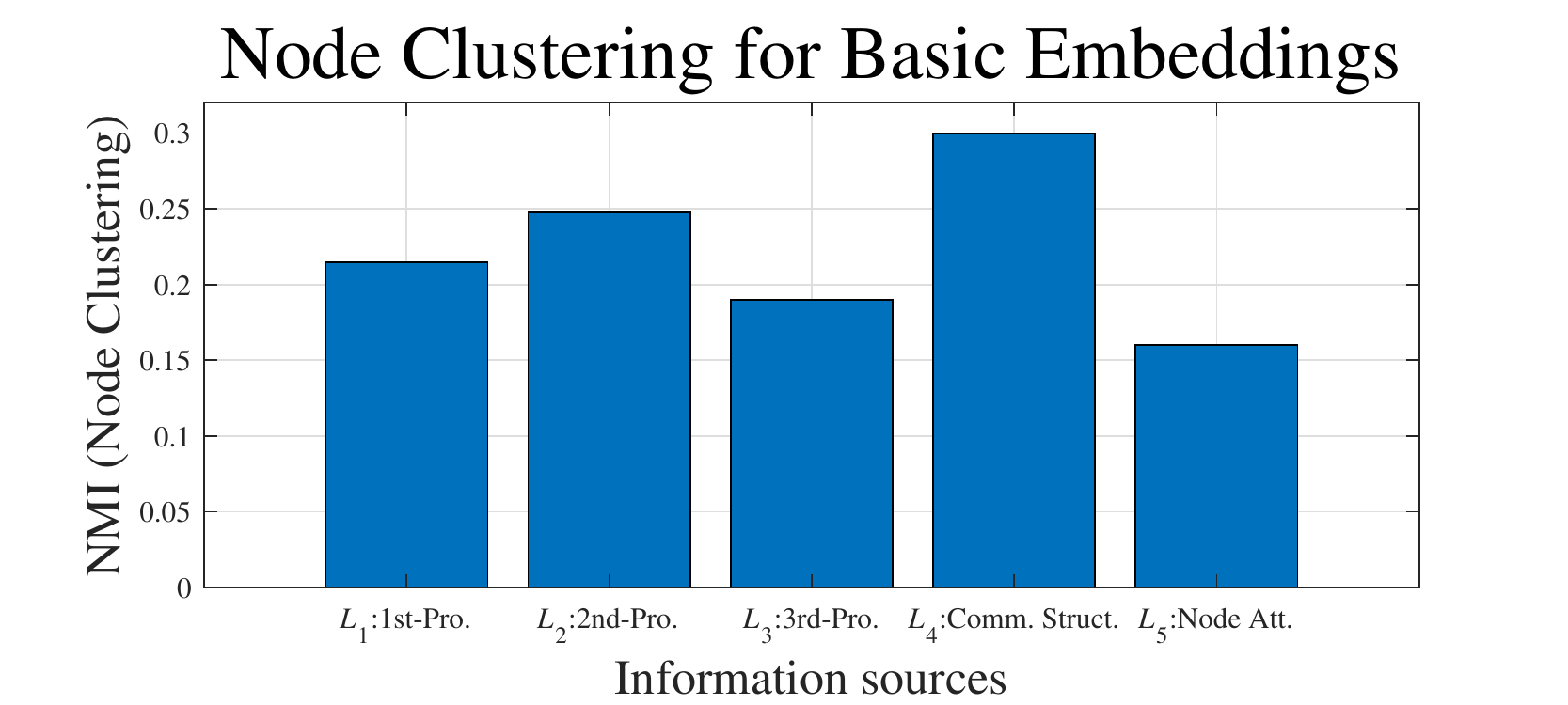}}
\end{minipage}
\begin{minipage}{0.24\linewidth}
    \centering
    \subfigure[AHGR{\tiny{(L)}} on \textit{TW}]
    {\includegraphics[width=1.0\textwidth, trim=20 0 45 5, clip]{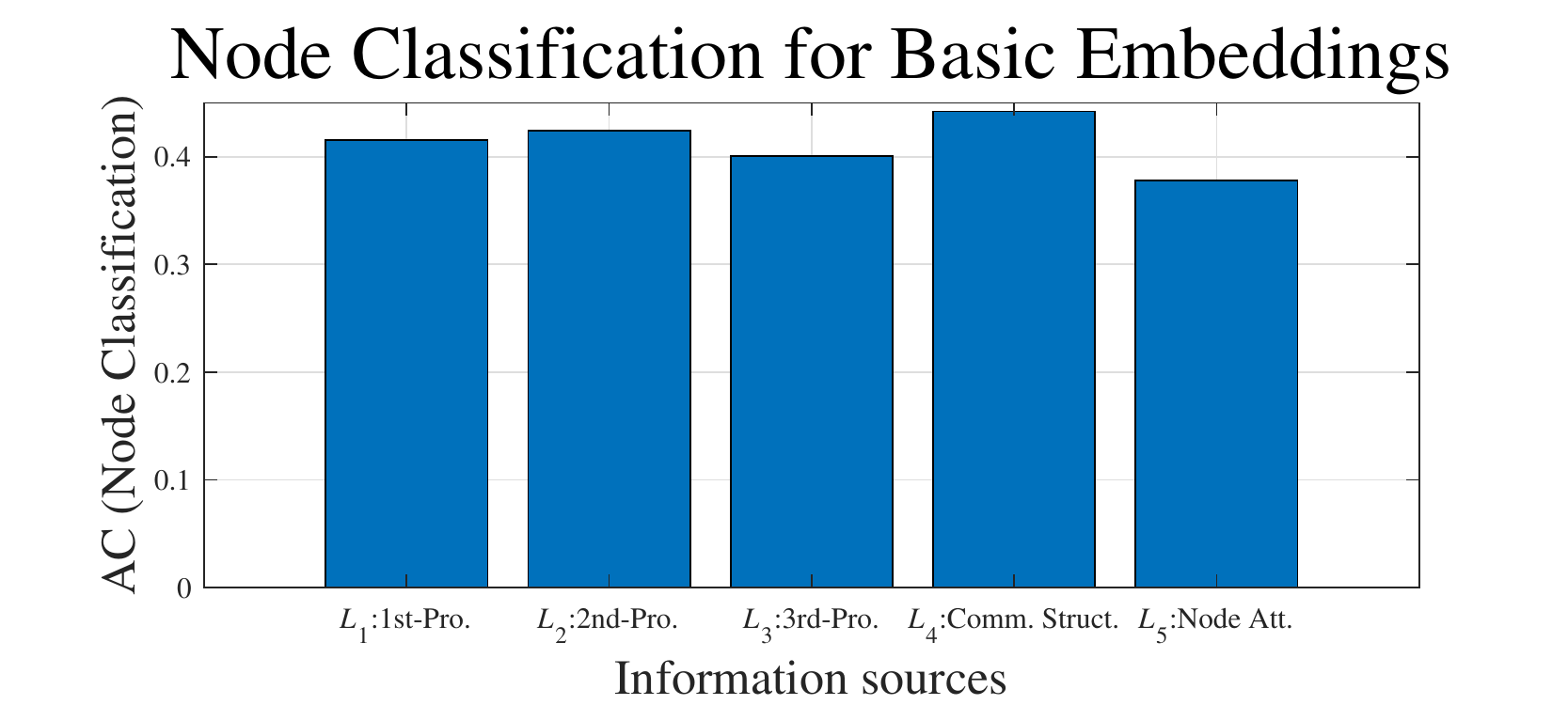}}
\end{minipage}
\caption{Analysis of the consistency indicator's effect with the indicator values ((a), (d), (g) and (j)) and the basic embedding performances for node clustering ((b), (e), (h) and (k)) as well as classification ((c), (f), (i) and (l))}
\label{Inds-Vis}
\vspace{-0.5cm}
\end{figure*}

To further validate the effect of \textit{consistency indicator}, we recorded values of indicators and performance of basic embeddings for both node clustering and classification. We illustrate two examples of \textit{Wisconsin} and \textit{Twitter} in Fig.~\ref{Inds-Vis}, where (a), (d), (g) and (j) are values of \textit{consistency indicators}; (b), (e), (h) and (k) are performance of node clustering (in terms of NMI); (c), (f), (i) and (l) are performance of node classification (in terms of AC).

All the results in Fig.~\ref{Inds-Vis} have consistent tendencies. The information sources with high indicator values have high quality metrics w.r.t. their basic embeddings for both node clustering and classification. On \textit{Wisconsin} (see Fig.~\ref{Inds-Vis} (a)-(f)), node attributes' higher indicator value corresponds to the better performance of its basic embedding for AHGR{\scriptsize{(N)}} and AHGR{\scriptsize{(L)}}. It implies a reasonable semi-supervised manner for the adjustment of parameters $\{ {\delta _T},{\delta _C},{\delta _A}\}$). Concretely, we can first evaluate the quality of each source's basic embedding by using only a small fraction of the application's ground-truth (e.g., labels of node clustering). The evaluation result can effectively direct the parameter setting w.r.t. the downstream application. For instance, the information ${I_l}$ with higher (lower) basic embedding quality should have more (less) contribution in the hybrid model, corresponding to the lower (higher) value of ${\delta_l}$.


\section{Conclusion}\label{Con}
In this paper, we adopted a graph reweighting scheme to formulate the graph representation learning with multiple information sources and proposed a novel AHGR method, which can potentially integrate arbitrary available sources. Since we introduced an NMF-based transition relation, AHGR can effectively perceive and resist the possible \textit{inconsistency} among different information sources. In addition, we also derived a novel \textit{consistency indicator} that can quantitatively measure a certain source's \textit{inconsistency degree}. Extensive experiments on synthetic and real graphs further validated the robustness of AHGR and effect of \textit{consistency indicator}.

In our future work, we will consider the integration of other available information (e.g., motif \cite{Benson2016Higher}, temporal topology \cite{lei2018adaptive,lei2019gcn,qin2022temporal,qin2023high}, etc). Moreover, we intend to further reduce the computation time of AHGR via the distributed implementation and optimized matrix operation libraries. To explore a semi-supervised parameter setting strategy based on the observation in Fig.~\ref{Inds-Vis} is also our next focus.


\bibliographystyle{IEEEtran}

\end{document}